  \documentclass[twocolumn,showpacs,preprintnumbers,amsmath,amssymb]{revtex4-2}

\usepackage[colorlinks=true,citecolor=blue,urlcolor=blue,linkcolor=blue,bookmarksopen]{hyperref}
\usepackage{amsfonts}
\usepackage{amsmath}
\usepackage{amssymb}
\usepackage{bbold}
\usepackage{mathrsfs}
\usepackage{graphicx}
\usepackage{dcolumn}
\usepackage{bm}
\usepackage{times}
\usepackage{comment}
\usepackage{latexsym}
\usepackage{epsf}
\usepackage{float}
\usepackage{natbib}
\usepackage{bm}
\usepackage[export]{adjustbox}
\usepackage[makeroom]{cancel}
\usepackage[english]{babel}
\usepackage[T1]{fontenc}
\usepackage{wasysym}
\usepackage[tight, FIGTOPCAP, hang, raggedright, nooneline]{subfigure}
\usepackage[dvipsnames]{xcolor}
\usepackage{epsfig}
\usepackage{lipsum}
\usepackage{enumitem}
\usepackage{mathtools}

\usepackage{ragged2e}

\DeclarePairedDelimiter\ket{\lvert}{\rangle}
\DeclarePairedDelimiterX\braket[2]{\langle}{\rangle}{#1 \delimsize\vert #2}
\renewcommand{\vec}[1]{\boldsymbol{#1}}

\begin{document}

\title{Band Inversion Flips the Winding of Bound States in the Continuum}

\author{Paul Bouteyre$^1$}
    \email{p.bouteyre@sheffield.ac.uk}
\author{Dung Xuan Nguyen$^{2,3,4}$}
\email{dungmuop@gmail.com}
\author{Loïc Malgrey$^{1,5}$}
\author{Zhiyi Yuan$^{6,7}$}
\author{Guillaume Gachon$^1$}
\author{Taha Benyattou$^1$}
\author{Xavier Letartre$^1$}
\author{Pierre Viktorovitch$^1$}
\author{S\'egol\`ene Callard$^1$}
\author{Guangwei Hu$^{6,7}$}
\author{Shanhui Fan$^8$}
\author{Lydie Ferrier$^1$}
    \email{lydie.ferrier@insa-lyon.fr}
\author{Hai Son Nguyen$^{1,7}$}
    \email{hai-son.nguyen@ec-lyon.fr}
    
\affiliation{$^1$Ecole Centrale de Lyon, CNRS, INSA Lyon, Universit\'e Claude Bernard Lyon 1, CPE Lyon, CNRS, INL, UMR5270, 69130 Ecully, France} 
\affiliation{$^2$Brown Theoretical Physics Center and Department of Physics, Brown University, 182 Hope Street, Providence, Rhode Island 02912, USA}   \affiliation{$^3$Center for Theoretical Physics of Complex Systems, Institute for Basic Science(IBS), Daejeon, Korea, 34126} 
 \affiliation{$^4$Institute for Interdisciplinary Research in Science and Education,
ICISE, Quy Nhon, 55131, Vietnam.} 

\affiliation{$^5$ Département de Physique, Ecole Normale Supérieure de Lyon,  46 Allée d’Italie, F 69342, Cedex 07 Lyon, France}
\affiliation{$^6$ School of Electrical and Electronic Engineering, Nanyang Technological University, Singapore 639798
}
\affiliation{$^7$
CNRS-International-NTU-Thales Research Alliance (CINTRA), IRL 3288, Singapore }

\affiliation{$^8$Ginzton Laboratory and Department of Electrical Engineering, Stanford University, Stanford, CA 94305, USA}
\date{\today}	

\begin{abstract}
Bound states in the continuum (BICs) in photonic slabs and metasurfaces appear as polarization singularities in momentum space, characterized by an integer winding number. This winding is widely treated as a robust topological label, preserved under smooth deformations of the structure. Here we show that this robustness fails under band inversion. Using a general two-band theory of open periodic photonic structures, we prove that a band inversion at a band-edge BIC reverses the local far-field polarization map and flips the BIC winding, $w_{\rm BIC}\to -w_{\rm BIC}$, \emph{without} any defect dynamics in momentum space. We verify the prediction in a tunable subwavelength grating, where polarization tomography directly images the reversal, and confirm it numerically in multiband rectangular and triangular photonic lattices. Band inversion thus emerges as a key mechanism governing polarization-singularity topology in non-Hermitian photonic band structures.
\end{abstract}

\maketitle
\emph{Introduction---}
Bound states in the continuum (BICs) are non-radiative eigenstates embedded in a continuum of radiative modes, enabled either by symmetry protection or by destructive interference between leakage channels~\cite{Friedrich1985,Hsu2016,Yin20,Joseph2021,Xu2023}. In photonic crystal slabs and metasurfaces, BICs appear as singular points of the far-field polarization field in momentum space, around which the polarization vector winds by an integer number $w$, providing a natural topological label that is widely viewed as robust under smooth structural deformations~\cite{Zhen2014,Yoda,Yin20}. A central consequence is that BICs behave as topological defects whose number, position, and charge can be reorganized by tuning structural parameters --- through pair creation/annihilation, merging, and splitting~\cite{Zhen2014,Jin2019,Liu2019,Yoda,Yin2020,Kang2021,Kang2022,Liu2024,Xu2025,Bai2025,Kang2025,Cui2025}. This hierarchy provides a flexible language for engineering high-$Q$ resonances~\cite{Contractor2022,Ren2022,Le2024,Do2025,Zhou2023} and shaping far-field radiation~\cite{Yin2020,Kang2025}.

In open photonic structures, radiation leakage renders the band structure intrinsically \emph{non-Hermitian}: eigenfrequencies become complex and the eigenvectors encode both modal composition and radiative coupling. Recent advances have therefore developed a non-Hermitian band-structure viewpoint for metasurfaces and photonic-crystal slabs, where loss and channel interference govern not only linewidths but also polarization textures and singularities~\cite{Letartre2022,Yuan2025,Lee2025,nguyen_generalized_2025}. Within this framework, a BIC can be regarded as the zero-leakage limit of a resonance whose far-field radiation amplitudes vanish. A central question then arises: \emph{which topological aspects of BIC polarization vortices remain robust when the band structure itself is engineered in the presence of radiative loss?}

In this Letter we show that \emph{band inversion alone --- with no defect motion in momentum space --- flips the winding number of a band-edge BIC.} The mechanism is the eigenvector exchange that defines a band inversion: as two hybridized branches swap their modal character across a gap closing, the BIC is transferred between them and the local linearized far-field polarization map is reversed, enforcing $w_{\rm BIC}\to -w_{\rm BIC}$ for both symmetry-protected $\Gamma$-point BICs and band-edge Friedrich--Wintgen BICs~\cite{Joseph2021,MermetLyaudoz2023,Le2024}. We verify this prediction in a tunable subwavelength grating, where angle-resolved spectroscopy reveals the inversion of a non-radiative \emph{dark} branch hosting a $\Gamma$-point BIC with its radiative \emph{bright} partner, and polarization-resolved tomography directly images the reversal of the BIC vortex circulation across the transition. By full-wave simulations of rectangular and triangular photonic-crystal slabs, we further show that the rule remains valid in multiband non-Hermitian band structures across multiple dark/bright permutation pathways, including configurations where symmetry-protected and Friedrich--Wintgen BICs coexist on a single branch. The winding number is therefore conserved under smooth deformations that preserve band ordering, but is \emph{not} an inversion-stable topological label.

\begin{figure}[ht!]
\begin{center}
\includegraphics[width=\linewidth]{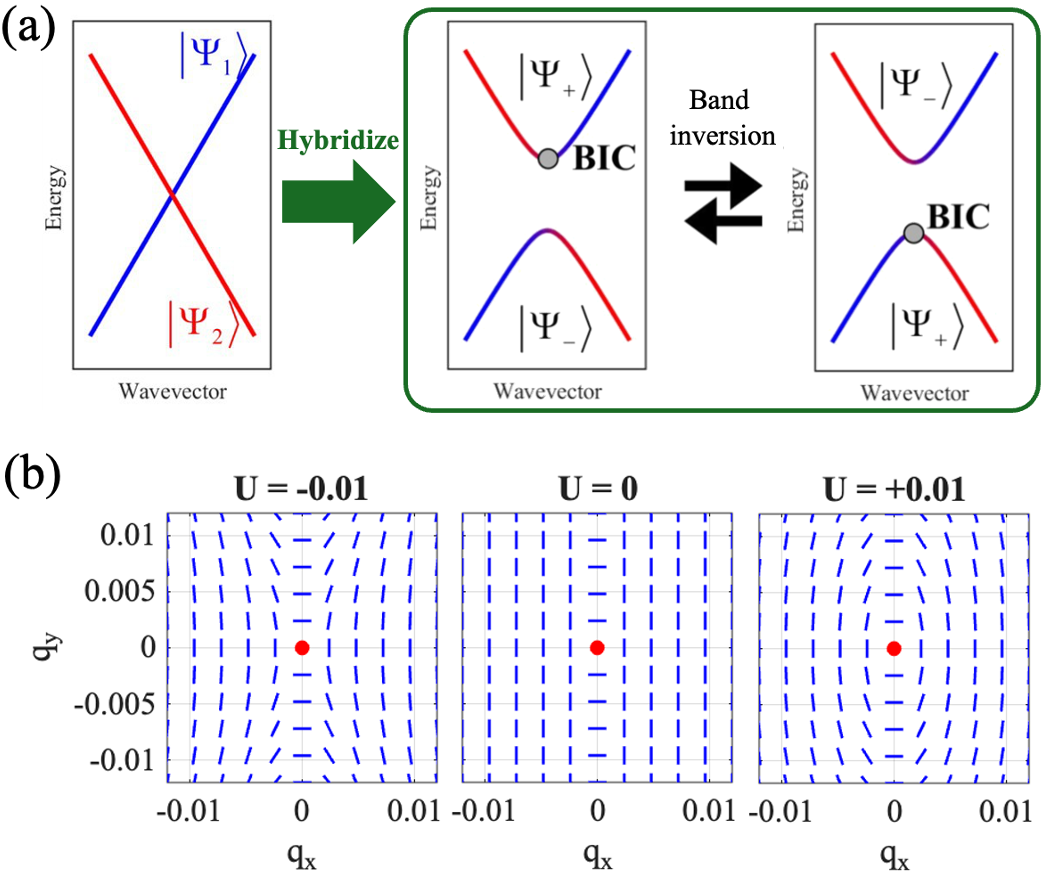}
\caption{\textbf{Mechanism and minimal model of the band-inversion-induced winding flip.}
\textbf{(a)} Hybridization of two lattice resonances $|\Psi_1\rangle$ and $|\Psi_2\rangle$ gives rise to two avoided-crossing branches $|\Psi_\pm\rangle$ with a BIC formed near the band edge of $|\Psi_+\rangle$. A band inversion corresponds to an exchange of modal character across a gap closing: the branch that is predominantly $|\Psi_1\rangle$-like on one side of the transition becomes $|\Psi_2\rangle$-like on the other side.
\textbf{(b)} Headless vector-field texture of the asymptotic far-field polarization near the $\Gamma$-point BIC for three values of the inversion parameter $U$, obtained from the effective non-Hermitian grating Hamiltonian with $n_g=8$ and $\gamma=0.001$. The plotted field is the dark-branch far field near $\Gamma$, with components $E_x\propto -2q_y$ and $E_y\propto \frac{q_x}{n_g}\frac{U-i\gamma}{U^2+\gamma^2}$ [Eq.~\eqref{eq:grating_farfield}]. For $U<0$ the polarization texture forms a first-order vortex; at $U=0$ it collapses into a locally shear-like, rank-deficient pattern with the winding undefined; for $U>0$ the vortex re-forms with opposite handedness.}
\label{fig:Bandinversion}
\end{center}
\end{figure}

\emph{Theory of BICs under band inversion---}
We develop a general two-band description of non-Hermitian band inversion in open periodic photonic structures. The treatment is platform-agnostic and applies to a broad class of metasurfaces and photonic-crystal slabs, where band inversions can be induced by tuning geometric or material parameters (e.g., slab thickness, filling fraction, refractive-index contrast), which modify near-field hybridization and/or radiative leakage.
Here we consider two lattice resonances $|\Psi_1\rangle$ and $|\Psi_2\rangle$ centered around Bloch wavevectors $\vec{K}_1^{0}$ and $\vec{K}_2^{0}$, with in-plane momenta $\vec{K}_m=\vec{K}_m^0+\vec{k}$,
where $\vec{k}$ is the wavevector in the reduced Brillouin zone. The microscopic nature of the two modes is not essential; we only assume that they (i) hybridize through near-field/lattice interaction and (ii) exchange radiative loss through their common coupling to the external continuum. These two ingredients are generic in open photonic lattices and produce an avoided crossing with complex eigenfrequencies, where one branch can exhibit strongly suppressed radiation at (or near) a band edge, forming a BIC (see Fig.~\ref{fig:Bandinversion}). The BIC considered here can be either symmetry-protected at the $\Gamma$-point band edge or interference-induced (Friedrich--Wintgen) and pinned at an off-$\Gamma$ band edge.


Near the BIC, the relevant eigenmode can be written as
\begin{equation}
|\Psi_+(\vec{k})\rangle=\alpha_1(\vec{k})|\Psi_1(\vec{k})\rangle+\alpha_2(\vec{k})|\Psi_2(\vec{k})\rangle,
\label{eq:eig_superposition}
\end{equation}
with complex coefficients $\alpha_m(\vec{k})$ defined up to an overall gauge. Because the structure is open, each basis mode radiates into a set of open channels $c$ (diffraction orders, up/down ports, polarizations). Denote by $D_{cm}(\vec{k})$ the complex radiation amplitude from basis mode $m$ into channel $c$. In experiments and in the polarization maps used to define vortices, one typically analyzes a \emph{collected} (projected) field, e.g. the specular order within a given numerical aperture and output side. We therefore introduce a linear projection $\mathcal P$ onto the collected subspace and define the corresponding measured Jones vector
\begin{equation}
\begin{aligned}
    \mathbf E(\vec{k})\equiv \mathcal P\,\mathbf E_{\rm out}(\vec{k})
&=\sum_{m=1,2}\alpha_m(\vec{k})\,\mathbf d_m(\vec{k})\\
\mathbf d_m(\vec{k})&\equiv \mathcal P\,\mathbf D_m(\vec{k}),
\end{aligned}
\label{eq:measured_field}
\end{equation}
where $\mathbf D_m(\vec{k})=(D_{cm}(\vec{k}))_{c\in{\rm open}}$ collects all channel amplitudes and $\mathbf d_m(\vec{k})$ is the projected coupling vector. Equation~\eqref{eq:measured_field} is general: it includes multiple diffraction orders and both polarizations, and reduces to the usual single-channel description when $\mathcal P$ selects a single output channel.  We rewrite this as $\mathbf E(\vec{k})=\alpha_2(\vec{k})[r(\vec{k})\,\mathbf d_1(\vec{k})+\mathbf d_2(\vec{k})]$, with mixing ratio $r(\vec{k})\equiv\alpha_1(\vec{k})/\alpha_2(\vec{k})$. The BIC condition $\mathbf E(\vec{k}^{\rm BIC})=\mathbf 0$ forces collinearity, $\mathbf d_2^{\rm BIC}=-r^{\rm BIC}\mathbf d_1^{\rm BIC}$, so the residual gauge freedom on the basis-mode phases lets us take $\mathbf d_{1,2}^{\rm BIC}\in\mathbb R$ and hence $r^{\rm BIC}\in\mathbb R$, with $\cos\beta^{\rm BIC}=\pm 1$ (see SM).

Since $\mathbf E(\vec{k}^{\rm BIC})=\mathbf 0$, the far-field polarization orientation angle $\phi$ is ill-defined at the BIC, and the associated topological charge (winding number) is~\cite{Zhen2014}
$w_{\rm BIC}=\frac{1}{2\pi}\oint_{\mathcal C}\nabla_{\vec{k}}\phi(\vec{k})\cdot d\vec{k}$, where $\mathcal C$ is a small loop enclosing $\vec{k}^{\rm BIC}$.
To evaluate the winding, it suffices to retain the leading (linear) expansion of the collected far field around the BIC.
We introduce a local polarization basis $\{\hat{\mathbf e}_{\parallel},\hat{\mathbf e}_{\perp}\}$ fixed at the BIC, with
$\hat{\mathbf e}_{\parallel}\parallel \mathbf d_1^{\rm BIC}$ and $\hat{\mathbf e}_{\perp}\perp \hat{\mathbf e}_{\parallel}$, and decompose
$\mathbf E=E_{\parallel}\hat{\mathbf e}_{\parallel}+E_{\perp}\hat{\mathbf e}_{\perp}$.
In this basis, the gauge-invariant polarization-ellipse orientation is defined via the Stokes parameters $S_1(\vec{k})=|E_{\parallel}(\vec{k})|^2-|E_{\perp}(\vec{k})|^2$ and $S_2(\vec{k})=2\,\Re\!\big(E_{\parallel}(\vec{k})E_{\perp}^*(\vec{k})\big)$ as $\phi(\vec{k})=\tfrac{1}{2}\arg\!\big(S_1(\vec{k})+iS_2(\vec{k})\big)$ (mod $\pi$)~\cite{DENNIS2002}. In the gauge we adopt at the BIC and to leading order in $\delta\vec{k}$, the linearized $E_{\parallel,\perp}$ are real (small-chirality regime, $|S_3|\!\ll\!|S_1|,|S_2|$), and this expression reduces to $\phi(\vec{k})=\arg\!\big(E_{\parallel}(\vec{k})+iE_{\perp}(\vec{k})\big)$ (mod $\pi$).
A direct linearization of the general two-mode radiation map (see Supplemental Material) yields the generic band-edge forms
\begin{align}
    E_{\perp}(\vec{k}^{\rm BIC}+\delta\vec{k})
&=\mathbf g_{\perp}\cdot \delta\vec{k}+O(|\delta\vec{k}|^2),
\label{eq:Eperp_linear}\\
E_{\parallel}(\vec{k}^{\rm BIC}+\delta\vec{k})
&=\left[A\,(\nabla_{\vec{k}} r)_{\rm BIC} + \mathbf g_{\parallel}\right]\cdot\delta\vec{k}
+O(|\delta\vec{k}|^2)
\label{eq:Epar_linear_simple}
\end{align}
with $A=\alpha_2^{\rm BIC}d_{1,\parallel}^{\rm BIC}$, 
$\mathbf g_{\perp}=\alpha_2^{\rm BIC}\Big(\nabla_{\vec{k}}[\,r^{\rm BIC}d_{1,\perp}+d_{2,\perp}\,]\Big)_{\rm BIC}$, and
$\mathbf g_{\parallel}=\alpha_2^{\rm BIC}\Big(\nabla_{\vec{k}}[\,r^{\rm BIC}d_{1,\parallel}+d_{2,\parallel}\,]\Big)_{\rm BIC}$.   Here the term $(\nabla_{\vec{k}} r)_{\rm BIC}$ in Eq.~\eqref{eq:Epar_linear_simple} is dictated by the eigenvectors through the mixing ratio
$r(\vec{k})$, whereas $\mathbf g_{\parallel}$ and $\mathbf g_{\perp}$ in Eqs.~\eqref{eq:Eperp_linear}, \eqref{eq:Epar_linear_simple} originate from the smooth $\vec{k}$-dependence of the projected radiation vectors $\mathbf d_{1,2}(\vec{k})$. In the gauge that we chose, $\mathbf g_{\parallel}$ and $\mathbf g_{\perp}$ are real.

In the two-mode description, each hybridized branch is characterized by the mixing ratio
$r(\vec{k})$, which encodes the relative weight of the two basis modes.
A band inversion corresponds to an exchange of modal character across a gap closing:
the branch that is predominantly $|\Psi_1\rangle$-like on one side of the transition becomes
$|\Psi_2\rangle$-like on the other side (see Fig.~\ref{fig:Bandinversion}). In the present open, non-Hermitian setting, the relevant gap closing is in the \emph{real part} of the complex eigenfrequencies; the imaginary parts (radiative losses) generically remain finite, so the two branches do not actually meet at a Hermitian-style degeneracy but rather encircle a pair of exceptional points (treated for the grating model in the Supplemental Material).
Locally at the BIC (pinned at the band edge), this exchange is reflected in a reversal of the
$\vec{k}$-dependence of the weight ratio,
\begin{equation}
(\nabla_{\vec{k}}|r|)_{\rm BIC}
\ \longrightarrow\
-\,(\nabla_{\vec{k}}|r|)_{\rm BIC},
\label{eq:inversion_signature_main}
\end{equation}
which signals that the slope of the modal weight across the band edge has changed sign. Mechanistically, near a gap closing the off-diagonal coupling between the two basis modes produces a slope $\partial_{\vec{k}}r\propto 1/\Delta$, where $\Delta$ is the gap separating the two branches; $\Delta$ flips sign through the inversion, hence so does $(\nabla_{\vec{k}}|r|)_{\rm BIC}$\,\footnote{In the open, non-Hermitian setting the relevant gap is complex, $\Delta=\Delta_{\rm Re}+i\,\Gamma_{\rm rad}$, with $\Gamma_{\rm rad}$ the radiative loss of the bright partner. The loss regularizes the magnitude $|\partial_{\vec k}r|$ to remain bounded by $\sim 1/(2\Gamma_{\rm rad})$ at the inversion but does not affect the parity: $\nabla_{\vec k}|r|^{\rm BIC}\propto\Delta_{\rm Re}/(\Delta_{\rm Re}^2+\Gamma_{\rm rad}^2)$ is odd in $\Delta_{\rm Re}$ and therefore still flips sign through the inversion. See the Supplemental Material for the full derivation and the Lorentzian-derivative profile of $\nabla_{\vec k}|r|^{\rm BIC}$ across the inversion.}.
By contrast, the projected radiation vectors $\mathbf g_{\parallel}$ and $\mathbf g_{\perp}$ arise from the smooth $\vec{k}$-dependence of the radiation channels and therefore vary continuously through the inversion.

We work in the regime where the radiation-channel amplitudes vary slowly with $\vec k$ on the scale of the BIC neighborhood, so that $\nabla_{\vec k}|\mathbf d_i|^{\rm BIC}=0$ to first order in $\delta\vec k$. Because $\hat{\mathbf e}_\parallel$ is parallel to $\mathbf d_i^{\rm BIC}$ (BIC collinearity), the projection $\hat{\mathbf e}_\parallel\cdot\nabla\mathbf d_i^{\rm BIC}=(\mathbf d_i\cdot\nabla\mathbf d_i)^{\rm BIC}/|\mathbf d_i^{\rm BIC}|$ vanishes identically; hence $\mathbf g_\parallel=0$ while $\mathbf g_\perp\neq 0$ (the variation of $\mathbf d_i$ near the BIC is purely transverse to $\hat{\mathbf e}_\parallel$). The more general regime where the channel magnitudes themselves vary is treated in the Supplemental Material. We then employ the standard expression for the polarization orientation in terms of the electric-field components~\cite{DENNIS2002}: $\tan(2\phi(\vec{k}))=\frac{S_2}{S_1}=\frac{2\,\Re(E_{\perp}^* E_\parallel)}{|E_{\perp}|^2-|E_\parallel|^2}$. Using Eqs.~\eqref{eq:Eperp_linear} and \eqref{eq:Epar_linear_simple} to evaluate $\tan(2\phi(\vec{k}))$, we have $\alpha_2^{\rm BIC}$ cancels out, so that the only remaining complex parameter is $r$. Near the BIC, $\cos(\beta(\vec{k})) = \pm 1 + O(|\delta \vec{k}|^2)$ so $\Re(\nabla_{\vec{k}}r)= \pm \nabla_{\vec{k}}|r|+ O(|\delta \vec{k}|^2)$. Under the band-inversion condition \eqref{eq:inversion_signature_main}, the numerator of $\tan(2\phi(\vec{k}))$ changes sign, which directly implies a sign reversal of the far-field polarization orientation.

Band inversion therefore reverses the polarization circulation and flips the BIC winding, $w_{\rm BIC}\to -w_{\rm BIC}$. This contrasts with Hermitian topological band theory, where band inversion redistributes Berry curvature while conserving the total Chern number~\cite{Avron:1983}: the relevant invariant there is a sum over bands. Here, by contrast, the invariant is carried by a single mode (the BIC) yet is still reversed by inversion.\\

 \emph{Geometric interpretation and the inversion transition---}
\emph{The winding number $w_{\rm BIC}$ is defined only as long as the BIC is an isolated first-order zero of the projected far field.}\,\footnote{Higher rotational symmetry ($C_4$, $C_6$) can forbid the linear transverse coupling, $\mathbf g_\perp=0$, giving a higher-order BIC with $|w_{\rm BIC}|\geq 2$; the leading polarization texture is then quadratic-or-higher in $\delta\vec k$. Band inversion either compensates through degenerate doublet partners (as for the $C_6$-symmetric triangular lattice in the End Matter) or splits the higher-order vortex into first-order ones once the rotational symmetry is broken~\cite{Yoda,Liu2019}.} When a structural parameter is tuned continuously through a band inversion, the measured far-field texture $\mathbf E(\vec{k})$ changes continuously --- so how can a discrete invariant flip sign? Writing $\vec{k}=\vec{k}^{\rm BIC}+\delta\vec{k}$, the leading-order expansion reads $E_{\parallel}=\mathbf a\!\cdot\!\delta\vec{k}$ and $E_{\perp}=\mathbf b\!\cdot\!\delta\vec{k}$, with $\mathbf a,\mathbf b\in\mathbb C^2$ (Eqs.~\eqref{eq:Eperp_linear}--\eqref{eq:Epar_linear_simple}). Since the polarization angle is $\phi=\arg(E_{\parallel}+iE_{\perp})$ (mod $\pi$), its winding is governed by the mapping $\delta\vec{k}\mapsto(\Re(E_{\parallel}+iE_{\perp}),\,\Im(E_{\parallel}+iE_{\perp}))$, which to leading order reads $(\Re(E_{\parallel}+iE_{\perp}),\,\Im(E_{\parallel}+iE_{\perp}))^{T}=J\,\delta\vec{k}+O(|\delta\vec{k}|^2)$ with $J=\big(\begin{smallmatrix}\Re(a_x+ib_x) & \Re(a_y+ib_y)\\ \Im(a_x+ib_x) & \Im(a_y+ib_y)\end{smallmatrix}\big)$. For an isolated first-order BIC, $\det J\neq 0$ and the winding equals the degree of this linear map, $w_{\rm BIC}=\mathrm{sgn}(\det J)$ (an equivalent statement is obtained from the Stokes trajectory $q(\theta)=S_1+iS_2$ in the small-chirality regime $|S_3|\ll|S_1|,|S_2|$ relevant for the open photonic band structures considered here, see Supplemental Material). Band inversion reverses the $\nabla_{\vec{k}}r$ contribution to $E_{\parallel}$, which flips $\det J$ and thus $w_{\rm BIC}\to -w_{\rm BIC}$. Because the tuning is continuous, $\det J$ must cross zero at the inversion point: $J$ becomes rank-deficient and the vortex is no longer first order, so the winding is \emph{undefined} exactly at the transition. Geometrically, the rotational texture collapses into a one-dimensional shear (locally parallel vectors) before re-forming with opposite handedness after inversion.
\begin{figure*}[hbt!]
\begin{center}
\includegraphics[width=0.8 \textwidth]{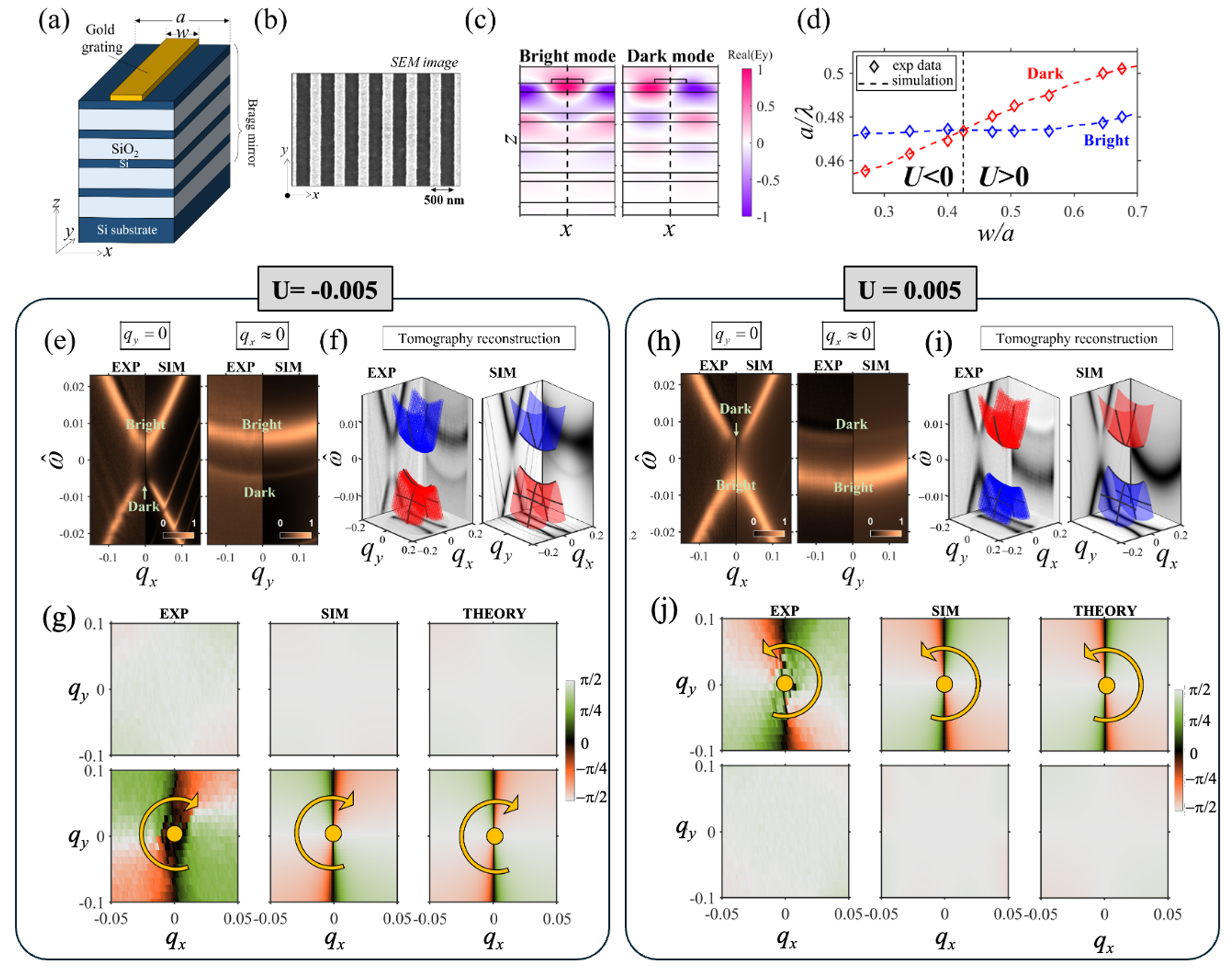}
\caption{\textbf{Experimental observation of band inversion and winding reversal in a 1D photonic lattice.}
(a,b) Sample schematic and SEM image of the fabricated gold grating on a dielectric Bragg mirror. 
(c) Field profiles of the two lowest TE modes at normal incidence, identifying the bright and dark branches. 
(d) Mode energies versus filling ratio $w/a$; the exchange of the dark and bright branches marks the band inversion. 
(e,h) Experimental and RCWA band structures for two samples on opposite sides of the transition ($U=-0.005$ and $U=+0.005$). 
(f,i) Corresponding reconstructed dispersion surfaces from experiment and simulation. 
(g,j) Momentum-space polarization-angle maps for the upper and lower branches, obtained from experiment, simulation, and theory. The dark branch hosts a polarization vortex whose circulation reverses across the inversion, while the bright branch remains vortex-free at $\Gamma$.\label{fig:grating_exp}}
\end{center}
\end{figure*}

\begin{figure}[ht!]
\includegraphics[width=\linewidth]{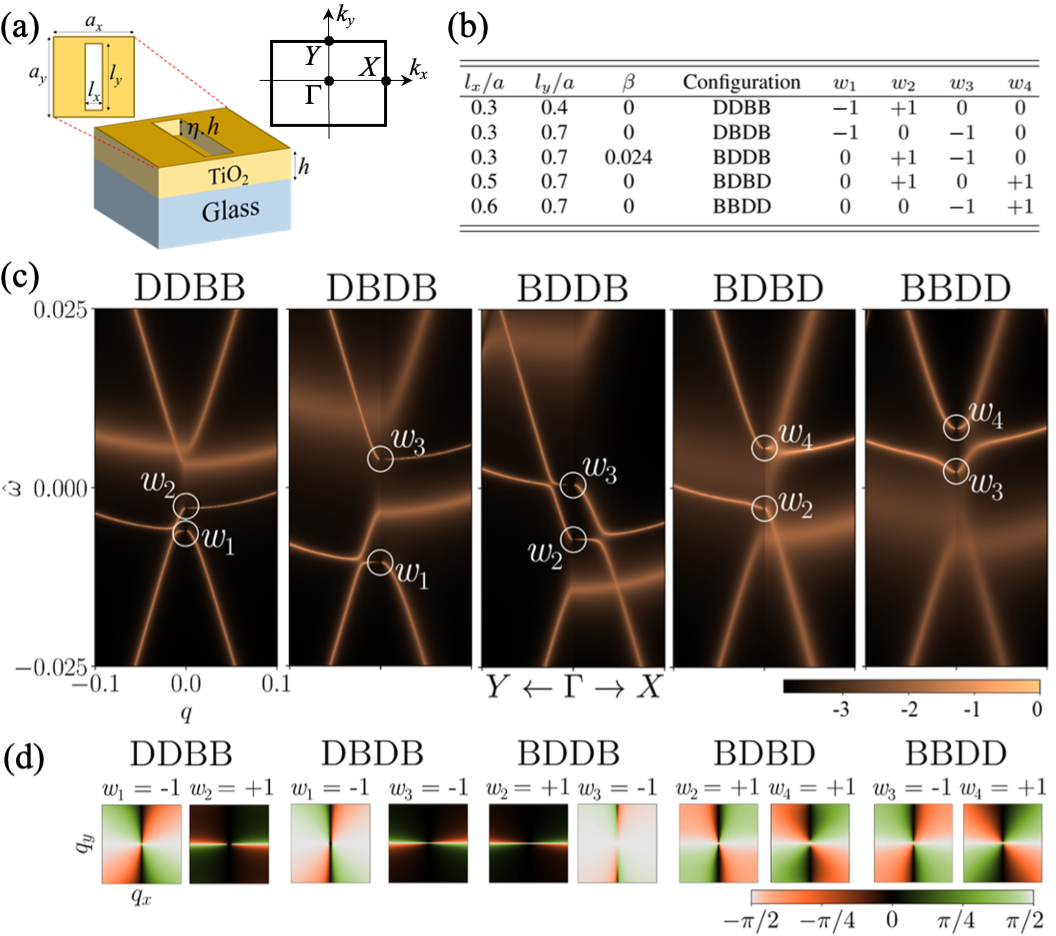}%
\caption{\textbf{Numerical demonstration of BIC winding reversal across band inversions.}
(a) Geometry of the simulated rectangular photonic-crystal slab: unit cell (left) and Brillouin zone (right).
(b) Band configurations near $\Gamma$ for five representative parameter sets. D (dark) labels a band hosting a symmetry-protected BIC at $\Gamma$; B (bright) labels a radiative band. Bands are ordered from lowest to highest energy, with the winding number $w_n$ of each BIC indicated.
(c) RCWA-computed angle-resolved absorption spectra (log scale), along $\Gamma Y$ ($q<0$) and $\Gamma X$ ($q>0$).
(d) Polarization-orientation textures in the $(q_x,q_y)$ plane, with momentum range $[-0.01,0.01]$.}
\label{fig:bandperm_2D}
\end{figure}

To illustrate this mechanism in a concrete physical setting, we consider a minimal model of the two lowest TE bands near $\Gamma$ in a 1D grating of period $a$ along the $x$-axis~\cite{Lee2019,Ferrier2022}. These bands originate from the hybridization of two counter-propagating guided waves and are described by the effective non-Hermitian Hamiltonian (see Supplemental Material for the derivation): $H=\left[\frac{q_y^2}{2n_g}-i(\gamma+\gamma_{nr})\right]\mathbb{I}_2+\left(U-i\gamma\cos\alpha\right)\sigma_1+\frac{q_x}{n_g}\sigma_3$, where $q_{x,y}=k_{x,y}a/2\pi$ are normalized momenta, $U$ is the diffractive coupling, $\gamma$ and $\gamma_{nr}$ are the radiative and non-radiative loss rates, and $n_g$ is the group index. The polarization-overlap factor is $\cos\alpha\simeq -1+q_y^2$ near $\Gamma$. At $\Gamma$ ($q_x=q_y=0$), the two eigenvalues are $\omega_\pm=\pm U - i(\gamma\mp\gamma+\gamma_{nr})$ and the corresponding eigenmodes $\ket{\pm}$ reduce to the antisymmetric and symmetric superpositions. The state $\ket{+}$ has vanishing radiative loss at $\Gamma$ and corresponds to the symmetry-protected BIC --- the dark branch --- while $\ket{-}$ is the bright branch. The sign of $U$ determines the ordering of these two branches, so that tuning $U$ through zero realizes a band inversion.

The dark branch carries the BIC vortex at $\Gamma$, and its projected far field near $\Gamma$ takes the asymptotic form (see Supplemental Material for the derivation)
\begin{equation}
E_x\propto -2q_y,\qquad
E_y\propto \frac{q_x}{n_g}\,\frac{U-i\gamma}{U^2+\gamma^2}.
\label{eq:grating_farfield}
\end{equation}
The prefactor of $q_x$ has $\Re\!\to\!-\Re$ across $U=0$, which is precisely the sign flip of $\nabla_{\vec{k}}|r|$ predicted by Eq.~\eqref{eq:inversion_signature_main} for this two-band model. Equation~\eqref{eq:grating_farfield} immediately shows how the vortex reverses across the inversion: the $q_y$-component is invariant while the $q_x$-component carries a prefactor whose real part changes sign with $U$, so the Jacobian of the local map from momentum space to far-field polarization reverses orientation through the transition. For $U\neq 0$ the BIC is an isolated first-order zero and the polarization field forms a vortex whose handedness is set by $\mathrm{sgn}(U)$. At the critical point $U=0$, however, the rotational texture collapses into a locally shear-like, quasi-one-dimensional pattern, so the first-order vortex structure is lost and the winding is undefined. This behavior is illustrated in Fig.~\ref{fig:Bandinversion}(b), where the vortex continuously unwinds at the inversion point and re-forms with opposite circulation on the other side of the transition.


\emph{Experimental demonstration in a two-band photonic lattice---}
We experimentally verify the band-inversion-induced flip of the BIC winding using the 1D grating platform discussed above, formed by a thin subwavelength gold grating on a dielectric Bragg mirror [Figs.~\ref{fig:grating_exp}(a,b)]. Near $\Gamma$, the two lowest TE modes have opposite parity under the mirror symmetry $x\rightarrow -x$ [Fig.~\ref{fig:grating_exp}(c)]: the antisymmetric mode is symmetry protected and forms the dark branch, while the symmetric mode is radiative and forms the bright branch.

The inversion parameter $U$ is tuned through the grating filling ratio $w/a$. Increasing $w/a$ shifts the dark mode strongly while leaving the bright mode nearly unchanged [Fig.~\ref{fig:grating_exp}(d)], exchanging their ordering and realizing a band inversion. We study two representative samples with $U=\pm 0.005$ by angle-resolved reflectivity-contrast spectroscopy. The measured band structures [Figs.~\ref{fig:grating_exp}(e,h)] and reconstructed dispersion surfaces [Figs.~\ref{fig:grating_exp}(f,i)] agree closely with rigorous coupled-wave analysis (RCWA) simulations~\cite{Liu2012} and directly evidence the dark/bright branch inversion. BICs are identified as local suppression of the guided-resonance feature. Polarization-resolved angle scans then reconstruct the momentum-space polarization texture of both branches [Figs.~\ref{fig:grating_exp}(g,j)]: the bright branch is vortex-free, while the dark branch hosts a vortex at $\Gamma$ whose circulation reverses across the inversion, in agreement with simulations and the Hamiltonian model. These measurements directly confirm that band inversion flips the winding number of a band-edge BIC.\\

\emph{Winding-number flips in multiband non-Hermitian band inversions---}
To demonstrate that the winding-flip rule remains valid beyond the minimal two-band picture, we consider a multiband 2D photonic-crystal slab consisting of a rectangular lattice of rectangular air holes etched into a TiO$_2$ slab on a glass substrate [Fig.~\ref{fig:bandperm_2D}(a)]. The lattice anisotropy is parametrized by $a_x=(1-\beta)a$ and $a_y=(1+\beta)a$ with $a=400$~nm, and the etching fraction is fixed to $\eta=0.5$ (50~nm etch depth in a total slab thickness $h=100$~nm). Lifting the $C_4$ symmetry of the square lattice down to $C_2$ resolves the degeneracy of the two bright modes at $\Gamma$ and allows individual dark/bright inversions to be tracked unambiguously (the $C_4$-symmetric case, where two compensating flips occur, is treated in the End Matter). Band inversions are induced by tuning $l_x$, $l_y$, and $\beta$ across the four lowest TE-like bands near $\Gamma$ --- two dark (D) branches hosting symmetry-protected BICs and two radiative bright (B) branches --- realizing several distinct D/B permutations in the same platform [Fig.~\ref{fig:bandperm_2D}(b)].

The band structures are extracted from angle-resolved absorption spectra computed by RCWA, with refractive indices $n_{\rm TiO_2}=2.4+10^{-6}i$ and $n_{\rm glass}=1.46$ [Fig.~\ref{fig:bandperm_2D}(c), along $\Gamma Y$ for $q<0$ and $\Gamma X$ for $q>0$, with $q_{x,y}$ normalized by $2\pi/a$]. For each configuration, the BIC is identified as the local disappearance of an otherwise well-defined resonance, and the far-field polarization-orientation texture is computed in the $(q_x,q_y)$ plane [Fig.~\ref{fig:bandperm_2D}(d)]. The results summarized in Fig.~\ref{fig:bandperm_2D}(b) confirm that, in every case where a dark and a bright branch exchange their ordering, the winding carried by the band-edge BIC flips sign, fully consistent with the general rule established above. The rule applies branch-by-branch and remains valid even when several BICs coexist on the same branch: a symmetry-protected $\Gamma$-point BIC together with two off-$\Gamma$ Friedrich--Wintgen BICs (see Fig.~\ref{fig:FWSP} in the End Matter, with $w_{\Gamma}+w_{+}+w_{-}=-1$ on the first band). The same principle holds in a six-band triangular-lattice photonic-crystal slab, treated in the End Matter.\\


\emph{Conclusion---} We have shown that band inversion in open non-Hermitian photonic lattices flips the winding of a band-edge BIC. A general two-band theory predicts that exchanging the modal character of two hybridized branches reverses the local polarization texture, so that the BIC winding is not an inversion-stable topological label. We verified the rule experimentally in a tunable 1D grating, where polarization tomography directly images the reversal of the vortex circulation, and confirmed it numerically in multiband rectangular and triangular photonic lattices, including configurations where symmetry-protected and Friedrich--Wintgen BICs coexist on the same branch. Beyond its conceptual implications, the rule turns band-inversion tuning into a deterministic recipe for reconfiguring polarization-vortex patterns in momentum space, with potential consequences for vortex-beam shaping and the on-demand spawning of circularly-polarized states~\cite{Liu2019}.\\

\emph{Acknowledgments---} The work is partly funded by the French National Research Agency (ANR) under the project POPEYE (ANR-17-CE24-0020) and the IDEXLYON from Universit\'e de Lyon, Scientific Breakthrough project TORE within the Programme Investissements d'Avenir (ANR-19-IDEX-0005). DXN was supported partly by Brown Theoretical Physics Center and by IBS-R024-D1. S.~F. acknowledges support from a MURI grant from the U.~S. Air Force Office of Scientific Research (AFOSR) (Grant No.~FA9550-21-1-0312).\\
\bibliography{main.bib}

@article{Joseph2021,
  title = {Bound states in the continuum in resonant nanostructures: an overview of engineered materials for tailored applications},
  volume = {10},
  ISSN = {2192-8614},
  url = {http://dx.doi.org/10.1515/nanoph-2021-0387},
  DOI = {10.1515/nanoph-2021-0387},
  number = {17},
  journal = {Nanophotonics},
  publisher = {Wiley},
  author = {Joseph,  Shereena and Pandey,  Saurabh and Sarkar,  Swagato and Joseph,  Joby},
  year = {2021},
  month = Nov,
  pages = {4175–4207}
}

@article{MermetLyaudoz2023,
  title = {Taming Friedrich–Wintgen Interference in a Resonant Metasurface: Vortex Laser Emitting at an On-Demand Tilted Angle},
  volume = {23},
  ISSN = {1530-6992},
  url = {http://dx.doi.org/10.1021/acs.nanolett.2c04936},
  DOI = {10.1021/acs.nanolett.2c04936},
  number = {10},
  journal = {Nano Letters},
  publisher = {American Chemical Society (ACS)},
  author = {Mermet-Lyaudoz,  Raphael and Symonds,  Clémentine and Berry,  Florian and Drouard,  Emmanuel and Chevalier,  Céline and Trippé-Allard,  Gaëlle and Deleporte,  Emmanuelle and Bellessa,  Joel and Seassal,  Christian and Nguyen,  Hai Son},
  year = {2023},
  month = May,
  pages = {4152–4159}
}

@article{Cui2025,
  title = {Ultracompact multibound-state-assisted flat-band lasers},
  volume = {19},
  ISSN = {1749-4893},
  url = {http://dx.doi.org/10.1038/s41566-025-01665-6},
  DOI = {10.1038/s41566-025-01665-6},
  number = {6},
  journal = {Nature Photonics},
  publisher = {Springer Science and Business Media LLC},
  author = {Cui,  Jieyuan and Han,  Song and Zhu,  Bofeng and Wang,  Chongwu and Chua,  Yunda and Wang,  Qian and Li,  Lianhe and Davies,  Alexander Giles and Linfield,  Edmund Harold and Wang,  Qi Jie},
  year = {2025},
  month = apr,
  pages = {643–649}
}

@article{Xu2023,
  title = {Recent Advances and Perspective of Photonic Bound States in the Continuum},
  volume = {3},
  ISSN = {2765-8791},
  url = {http://dx.doi.org/10.34133/ultrafastscience.0033},
  DOI = {10.34133/ultrafastscience.0033},
  journal = {Ultrafast Science},
  publisher = {American Association for the Advancement of Science (AAAS)},
  author = {Xu,  Guizhen and Xing,  Hongyang and Xue,  Zhanqiang and Lu,  Dan and Fan,  Jinying and Fan,  Junxing and Shum,  Perry Ping and Cong,  Longqing},
  year = {2023},
  month = jan 
}

@article{Liu2024,
  title = {Merging diverse bound states in the continuum: from intrinsic to extrinsic scenarios},
  volume = {32},
  ISSN = {1094-4087},
  url = {http://dx.doi.org/10.1364/OE.522480},
  DOI = {10.1364/oe.522480},
  number = {9},
  journal = {Optics Express},
  publisher = {Optica Publishing Group},
  author = {Liu,  Liangliang and Luo,  Haoqi and Lu,  Yonghua and Wang,  Pei},
  year = {2024},
  month = apr,
  pages = {16491}
}

@article{Xu2025,
  title = {Splitting and merging of bound states in the continuum through defect engineering},
  volume = {112},
  ISSN = {2469-9934},
  url = {http://dx.doi.org/10.1103/2w92-47q1},
  DOI = {10.1103/2w92-47q1},
  number = {3},
  journal = {Physical Review A},
  publisher = {American Physical Society (APS)},
  author = {Xu,  Ke and Liu,  Miao and Ren,  Hao and Deng,  Xuesong and Feng,  Jian and Fang,  Ming and Huang,  Zhixiang},
  year = {2025},
  month = sep 
}

@article{nguyen_generalized_2025,
  title={Generalized non‑Hermitian Hamiltonian for guided resonances in photonic crystal slabs},
  author={Nguyen, Viet Anh and Nguyen, Hung Son and Yuan, Zhiyi and Nguyen, Dung Xuan and Dang, Cuong and Ha, Son Tung and Letartre, Xavier and Le‑Van, Quynh and Nguyen, Hai Son},
  journal={Nanophotonics},
  volume={14},
  number={27},
  pages={5229--5250},
  year={2025},
  publisher={Walter de Gruyter},
  doi={10.1515/nanoph-2025-0393},
  url={https://doi.org/10.1515/nanoph-2025-0393}
}

@article{Ferrier2022,
  title = {Unveiling the Enhancement of Spontaneous Emission at Exceptional Points},
  volume = {129},
  ISSN = {1079-7114},
  url = {http://dx.doi.org/10.1103/PhysRevLett.129.083602},
  DOI = {10.1103/physrevlett.129.083602},
  number = {8},
  journal = {Physical Review Letters},
  publisher = {American Physical Society (APS)},
  author = {Ferrier,  L. and Bouteyre,  P. and Pick,  A. and Cueff,  S. and Dang,  N. H. M and Diederichs,  C. and Belarouci,  A. and Benyattou,  T. and Zhao,  J. X. and Su,  R. and Xing,  J. and Xiong,  Qihua and Nguyen,  H. S.},
  year = {2022},
  month = aug 
}

@article{Lee2025,
  title = {Dirac bilayer metasurfaces as an inverse Gires-Tournois etalon},
  volume = {7},
  ISSN = {2643-1564},
  url = {http://dx.doi.org/10.1103/3dw5-nvdq},
  DOI = {10.1103/3dw5-nvdq},
  number = {4},
  journal = {Physical Review Research},
  publisher = {American Physical Society (APS)},
  author = {Lee,  Ki Young and Yoo,  Kwang Wook and Monticone,  Francesco and Yoon,  Jae Woong},
  year = {2025},
  month = oct 
}

@article{Yuan2025,
  title = {Breakdown of bulk-radiation correspondence in radiative photonic lattices},
  volume = {7},
  ISSN = {2643-1564},
  url = {http://dx.doi.org/10.1103/1gbs-rdxm},
  DOI = {10.1103/1gbs-rdxm},
  number = {4},
  journal = {Physical Review Research},
  publisher = {American Physical Society (APS)},
  author = {Yuan,  Xinyi and Malgrey,  Loïc and Sigurðsson,  Helgi and Nguyen,  Hai Son and Salerno,  Grazia},
  year = {2025},
  month = nov 
}

@article{Letartre2022,
  title = {Analytical non-Hermitian description of photonic crystals with arbitrary lateral and transverse symmetry},
  volume = {106},
  ISSN = {2469-9934},
  url = {http://dx.doi.org/10.1103/PhysRevA.106.033510},
  DOI = {10.1103/physreva.106.033510},
  number = {3},
  journal = {Physical Review A},
  publisher = {American Physical Society (APS)},
  author = {Letartre,  Xavier and Mazauric,  Serge and Cueff,  Sébastien and Benyattou,  Taha and Nguyen,  Hai Son and Viktorovitch,  Pierre},
  year = {2022},
  month = sep 
}

@article{DENNIS2002,
title = {Polarization singularities in paraxial vector fields: morphology and statistics},
journal = {Optics Communications},
volume = {213},
number = {4},
pages = {201-221},
year = {2002},
issn = {0030-4018},
doi = {https://doi.org/10.1016/S0030-4018(02)02088-6},
url = {https://www.sciencedirect.com/science/article/pii/S0030401802020886},
author = {M.R. Dennis}
}

@article{Friedrich1985,
author = {Friedrich, H. and Wintgen, D.},
doi = {10.1103/PhysRevA.32.3231},
issn = {0556-2791},
journal = {Physical Review A},
month = {dec},
number = {6},
pages = {3231--3242},
title = {{Interfering resonances and bound states in the continuum}},
url = {https://link.aps.org/doi/10.1103/PhysRevA.32.3231},
volume = {32},
year = {1985}
}

@article{Zhen2014,
author = {Zhen, Bo and Hsu, Chia Wei and Lu, Ling and Stone, a. Douglas and Solja{\v{c}}i{\'{c}}, Marin},
doi = {10.1103/PhysRevLett.113.257401},
issn = {0031-9007},
journal = {Physical Review Letters},
number = {25},
pages = {1--5},
title = {{Topological Nature of Optical Bound States in the Continuum}},
url = {http://link.aps.org/doi/10.1103/PhysRevLett.113.257401},
volume = {113},
year = {2014}
}

@article{Yin20,
author = {Xuefan Yin and Chao Peng},
journal = {Photon. Res.},
number = {11},
pages = {B25--B38},
publisher = {OSA},
title = {Manipulating light radiation from a topological perspective},
volume = {8},
month = {Nov},
year = {2020},
url = {http://www.osapublishing.org/prj/abstract.cfm?URI=prj-8-11-B25},
doi = {10.1364/PRJ.403444},
}

@article{Lee2019,
  title = {Band flips and bound-state transitions in leaky-mode photonic lattices},
  author = {Lee, Sun-Goo and Magnusson, Robert},
  journal = {Phys. Rev. B},
  volume = {99},
  issue = {4},
  pages = {045304},
  numpages = {5},
  year = {2019},
  month = {Jan},
  publisher = {American Physical Society},
  doi = {10.1103/PhysRevB.99.045304},
  url = {https://link.aps.org/doi/10.1103/PhysRevB.99.045304}
}

@article{Yoda,
  title = {Generation and Annihilation of Topologically Protected Bound States in the Continuum and Circularly Polarized States by Symmetry Breaking},
  author = {Yoda, Taiki and Notomi, Masaya},
  journal = {Phys. Rev. Lett.},
  volume = {125},
  issue = {5},
  pages = {053902},
  numpages = {6},
  year = {2020},
  month = {Jul},
  publisher = {American Physical Society},
  doi = {10.1103/PhysRevLett.125.053902},
  url = {https://link.aps.org/doi/10.1103/PhysRevLett.125.053902}
}

@article{Hsu2016,
author = {Hsu, Chia Wei and Zhen, Bo and Stone, A. Douglas and Joannopoulos, John D. and Solja{\v{c}}i{\'{c}}, Marin},
doi = {10.1038/natrevmats.2016.48},
issn = {2058-8437},
journal = {Nature Reviews Materials},
number = {18},
pages = {16048},
pmid = {18518374},
title = {{Bound states in the continuum}},
url = {http://www.nature.com/articles/natrevmats201648},
volume = {1},
year = {2016}
}

@article{Jin2019,
author = {Jin, Jicheng and Yin, Xuefan and Ni, Liangfu and Solja{\v{c}}i{\'{c}}, Marin and Zhen, Bo and Peng, Chao},
doi = {10.1038/s41586-019-1664-7},
issn = {1476-4687},
journal = {Nature},
number = {7779},
pages = {501--504},
title = {{Topologically enabled ultrahigh-Q guided resonances robust to out-of-plane scattering}},
url = {https://doi.org/10.1038/s41586-019-1664-7},
volume = {574},
year = {2019}
}

@article{Kang2021,
  title = {Merging Bound States in the Continuum at Off-High Symmetry Points},
  author = {Kang, Meng and Zhang, Shunping and Xiao, Meng and Xu, Hongxing},
  journal = {Phys. Rev. Lett.},
  volume = {126},
  issue = {11},
  pages = {117402},
  numpages = {6},
  year = {2021},
  month = {Mar},
  publisher = {American Physical Society},
  doi = {10.1103/PhysRevLett.126.117402},
  url = {https://link.aps.org/doi/10.1103/PhysRevLett.126.117402}
}

@article{Bai2025,
  title = {Recovery of topologically robust merging bound states in the continuum in photonic structures with broken symmetry},
  volume = {14},
  ISSN = {2192-8614},
  url = {http://dx.doi.org/10.1515/nanoph-2024-0609},
  DOI = {10.1515/nanoph-2024-0609},
  number = {7},
  journal = {Nanophotonics},
  publisher = {Wiley},
  author = {Bai,  Huayu and Shevchenko,  Andriy and Kolkowski,  Radoslaw},
  year = {2025},
  month = mar,
  pages = {899–913}
}

@article{Yin2020,
  title = {Observation of topologically enabled unidirectional guided resonances},
  volume = {580},
  ISSN = {1476-4687},
  url = {http://dx.doi.org/10.1038/s41586-020-2181-4},
  DOI = {10.1038/s41586-020-2181-4},
  number = {7804},
  journal = {Nature},
  publisher = {Springer Science and Business Media LLC},
  author = {Yin,  Xuefan and Jin,  Jicheng and Soljačić,  Marin and Peng,  Chao and Zhen,  Bo},
  year = {2020},
  month = apr,
  pages = {467–471}
}

@article{Kang2025,
  title = {Janus Bound States in the Continuum with Asymmetric Topological Charges},
  volume = {134},
  ISSN = {1079-7114},
  url = {http://dx.doi.org/10.1103/PhysRevLett.134.013805},
  DOI = {10.1103/physrevlett.134.013805},
  number = {1},
  journal = {Physical Review Letters},
  publisher = {American Physical Society (APS)},
  author = {Kang,  Meng and Xiao,  Meng and Chan,  C. T.},
  year = {2025},
  month = jan 
}

@article{Contractor2022,
  title = {Scalable single-mode surface-emitting laser via open-Dirac singularities},
  volume = {608},
  ISSN = {1476-4687},
  url = {http://dx.doi.org/10.1038/s41586-022-05021-4},
  DOI = {10.1038/s41586-022-05021-4},
  number = {7924},
  journal = {Nature},
  publisher = {Springer Science and Business Media LLC},
  author = {Contractor,  Rushin and Noh,  Wanwoo and Redjem,  Walid and Qarony,  Wayesh and Martin,  Emma and Dhuey,  Scott and Schwartzberg,  Adam and Kanté,  Boubacar},
  year = {2022},
  month = jun,
  pages = {692–698}
}

@article{Zhou2023,
  title = {Increasing the Q-Contrast in Large Photonic Crystal Slab Resonators Using Bound-States-in-Continuum},
  volume = {10},
  ISSN = {2330-4022},
  url = {http://dx.doi.org/10.1021/acsphotonics.3c00126},
  DOI = {10.1021/acsphotonics.3c00126},
  number = {5},
  journal = {ACS Photonics},
  publisher = {American Chemical Society (ACS)},
  author = {Zhou,  Ming and Kumar Kalapala,  Akhil Raj and Pan,  Mingsen and Gibson,  Ricky and Reilly,  Kevin James and Rotter,  Thomas and Balakrishnan,  Garnesh and Bedford,  Robert and Zhou,  Weidong and Fan,  Shanhui},
  year = {2023},
  month = may,
  pages = {1519–1528}
}

@article{Le2024,
  title = {Super Bound States in the Continuum on a Photonic Flatband: Concept,  Experimental Realization,  and Optical Trapping Demonstration},
  volume = {132},
  ISSN = {1079-7114},
  url = {http://dx.doi.org/10.1103/PhysRevLett.132.173802},
  DOI = {10.1103/physrevlett.132.173802},
  number = {17},
  journal = {Physical Review Letters},
  publisher = {American Physical Society (APS)},
  author = {Le,  Ngoc Duc and Bouteyre,  Paul and Kheir-Aldine,  Ali and Dubois,  Florian and Cueff,  Sébastien and Berguiga,  Lotfi and Letartre,  Xavier and Viktorovitch,  Pierre and Benyattou,  Taha and Nguyen,  Hai Son},
  year = {2024},
  month = apr 
}

@article{Do2025,
  title = {Room-Temperature Lasing at Flatband Bound States in the Continuum},
  volume = {19},
  ISSN = {1936-086X},
  url = {http://dx.doi.org/10.1021/acsnano.5c01972},
  DOI = {10.1021/acsnano.5c01972},
  number = {20},
  journal = {ACS Nano},
  publisher = {American Chemical Society (ACS)},
  author = {Do,  Thi Thu Ha and Yuan,  Zhiyi and Durmusoglu,  Emek G. and Shamkhi,  Hadi K. and Valuckas,  Vytautas and Zhao,  Chunyu and Kuznetsov,  Arseniy I. and Demir,  Hilmi Volkan and Dang,  Cuong and Nguyen,  Hai Son and Ha,  Son Tung},
  year = {2025},
  month = may,
  pages = {19287–19296}
}

@article{Ren2022,
  title = {Low-threshold nanolasers based on miniaturized bound states in the continuum},
  volume = {8},
  ISSN = {2375-2548},
  url = {http://dx.doi.org/10.1126/sciadv.ade8817},
  DOI = {10.1126/sciadv.ade8817},
  number = {51},
  journal = {Science Advances},
  publisher = {American Association for the Advancement of Science (AAAS)},
  author = {Ren,  Yuhao and Li,  Peishen and Liu,  Zhuojun and Chen,  Zihao and Chen,  You-Ling and Peng,  Chao and Liu,  Jin},
  year = {2022},
  month = dec 
}

@Article{Kang2022,
  author   = {Kang, Meng and Mao, Li and Zhang, Shunping and Xiao, Meng and Xu, Hongxing and Chan, Che Ting},
  journal  = {Light: Science \& Applications},
  title    = {Merging bound states in the continuum by harnessing higher-order topological charges},
  year     = {2022},
  issn     = {2047-7538},
  number   = {1},
  pages    = {228},
  volume   = {11},
  doi      = {10.1038/s41377-022-00923-4},
  refid    = {Kang2022},
  url      = {https://doi.org/10.1038/s41377-022-00923-4},
}

@article{Liu2019,
  title = {Circularly Polarized States Spawning from Bound States in the Continuum},
  author = {Liu, Wenzhe and Wang, Bo and Zhang, Yiwen and Wang, Jiajun and Zhao, Maoxiong and Guan, Fang and Liu, Xiaohan and Shi, Lei and Zi, Jian},
  journal = {Phys. Rev. Lett.},
  volume = {123},
  issue = {11},
  pages = {116104},
  numpages = {6},
  year = {2019},
  month = {Sep},
  publisher = {American Physical Society},
  doi = {10.1103/PhysRevLett.123.116104},
  url = {https://link.aps.org/doi/10.1103/PhysRevLett.123.116104}
}

@article{Liu2012,
title = {S4 : A free electromagnetic solver for layered periodic structures},
journal = {Computer Physics Communications},
volume = {183},
number = {10},
pages = {2233-2244},
year = {2012},
issn = {0010-4655},
doi = {https://doi.org/10.1016/j.cpc.2012.04.026},
url = {https://www.sciencedirect.com/science/article/pii/S0010465512001658},
author = {Victor Liu and Shanhui Fan}
}

@article{Avron:1983,
  title = {Homotopy and Quantization in Condensed Matter Physics},
  author = {Avron, J. E. and Seiler, R. and Simon, B.},
  journal = {Phys. Rev. Lett.},
  volume = {51},
  issue = {1},
  pages = {51--53},
  numpages = {0},
  year = {1983},
  month = {Jul},
  publisher = {American Physical Society},
  doi = {10.1103/PhysRevLett.51.51},
  url = {https://link.aps.org/doi/10.1103/PhysRevLett.51.51}
}
\newpage
    \onecolumngrid
    \section*{End Matter}
    \twocolumngrid

\begin{figure}[hbt!]
\begin{center}
\includegraphics[width=\linewidth]{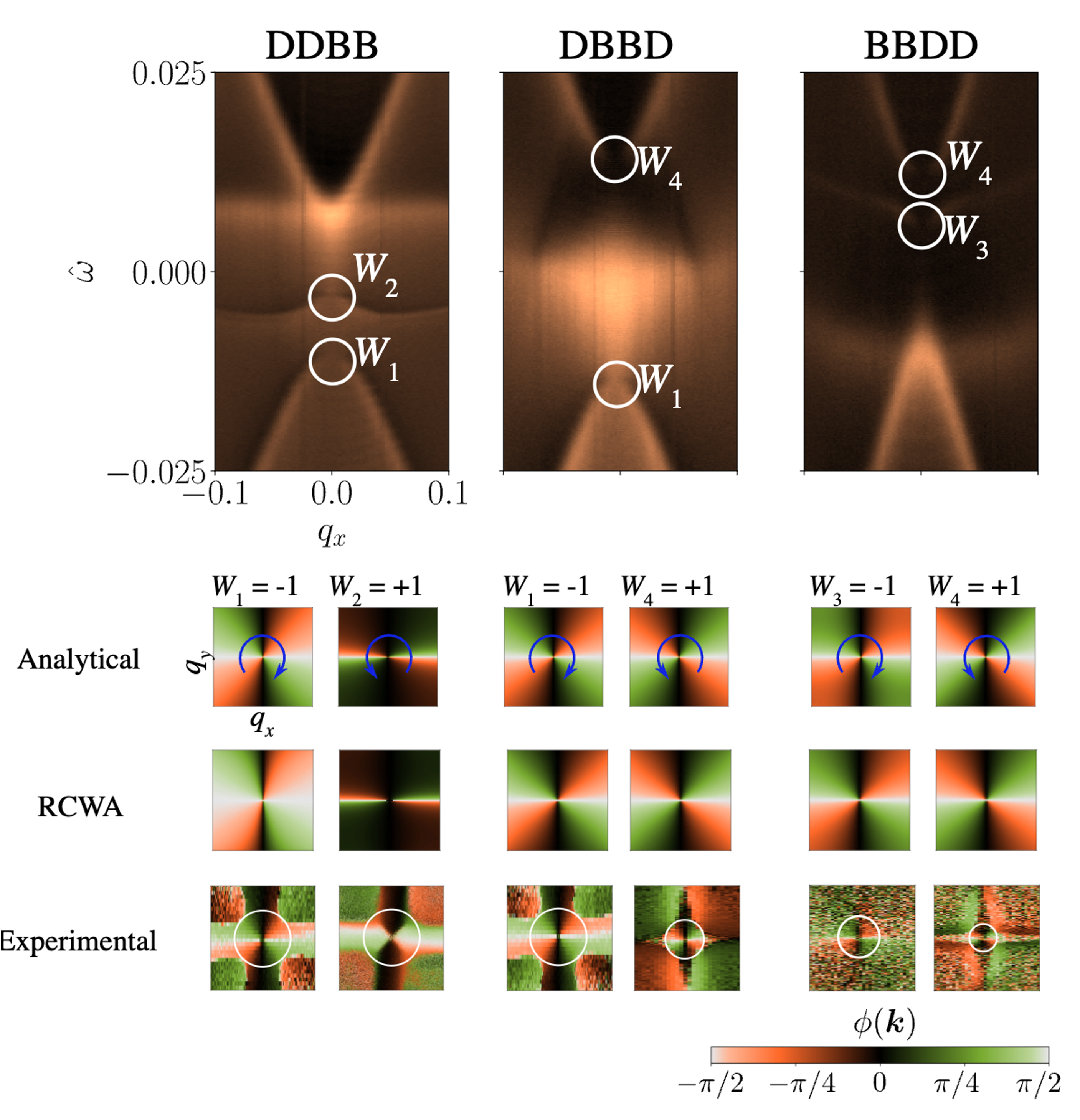}
\caption{\label{fig:C4_square_lattice}
\textbf{Experimental multiband band inversion in a $C_4$-symmetric square lattice.}
Top row: angle-resolved reflectivity spectra for three fabricated TiO$_2$ photonic-crystal slabs with square air holes, corresponding to the band orderings DDBB, DBBD, and BBDD as the hole-size ratio $l/a$ is varied. Because of $C_4$ symmetry, the two bright modes remain degenerate at $\Gamma$, while the two dark modes host symmetry-protected BICs. From one configuration to the next, the degenerate bright pair exchanges ordering successively with one dark mode, so that each dark branch undergoes two consecutive dark/bright inversions. Bottom rows: far-field polarization textures of the dark bands obtained from the effective non-Hermitian $4\times4$ model (Analytical), RCWA simulations, and experiment. Each individual inversion flips the winding of the corresponding band-edge BIC, but the two successive flips compensate, so the net winding of a given dark branch is unchanged between the DDBB and BBDD configurations. }
\end{center}
\end{figure}

\begin{figure}[hbt!]
\begin{center}
\includegraphics[width=\linewidth]{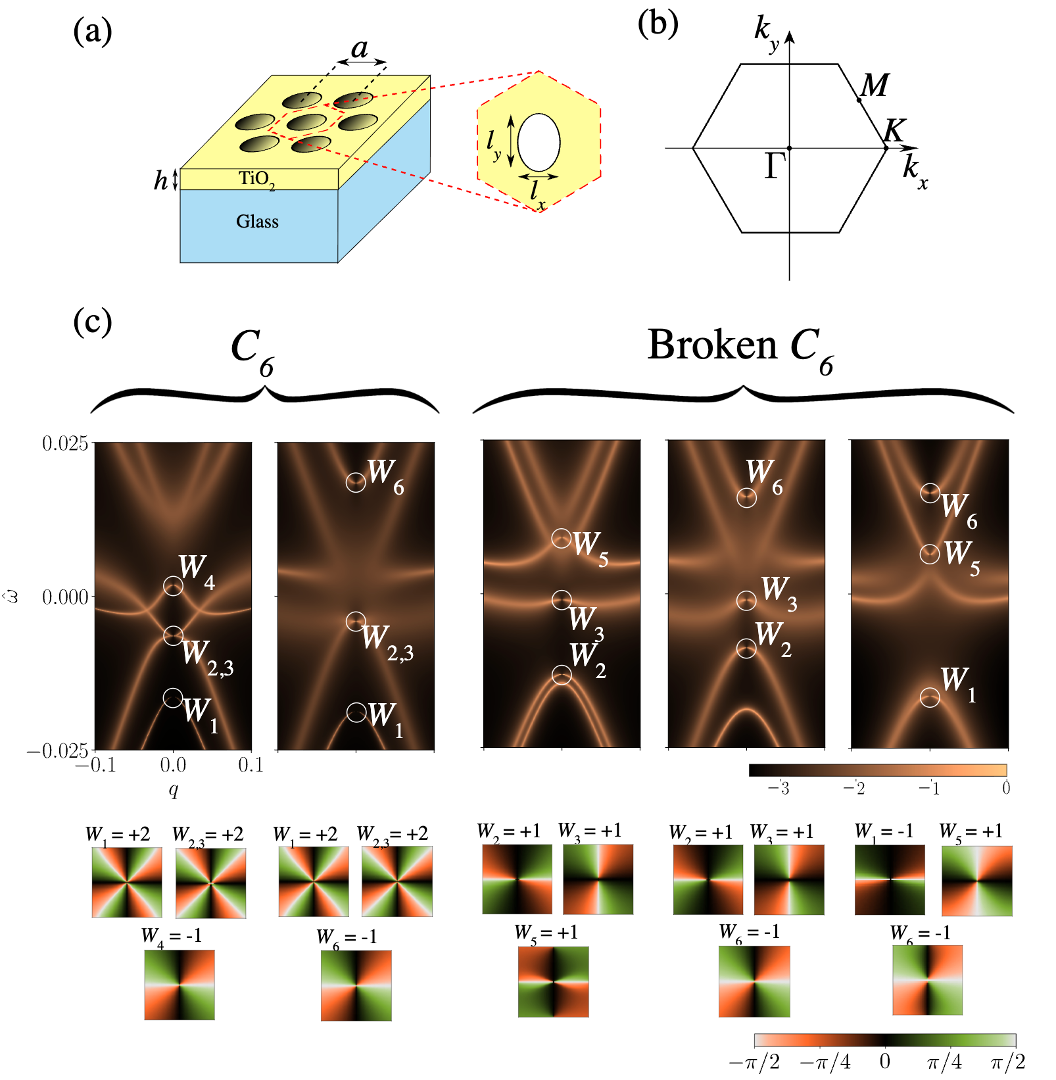}
\caption{\label{fig:triangular_lattice}
\textbf{Multiple band inversions in a triangular-lattice photonic-crystal slab.}
(a) Schematic of the simulated structure.
(b) First Brillouin zone of the triangular lattice. 
(c) RCWA absorption spectra near $\Gamma$ (top) and corresponding momentum-space polarization textures of the band-edge BICs (bottom) for representative configurations with preserved $C_6$ symmetry ($l_x=l_y$) and broken $C_6$ symmetry ($l_x\neq l_y$). In the $C_6$-symmetric case, the two bright modes remain degenerate, so that each dark branch effectively crosses a bright doublet and undergoes two successive winding flips that compensate, leaving the net winding unchanged. When $C_6$ is broken, all degeneracies are lifted while the number of dark and bright bands is preserved; the winding number then flips sign after each individual dark--bright band inversion. This example confirms the robustness of the winding-flip rule beyond the minimal two-band setting.}
\end{center}
\end{figure}

\subsection*{Why the winding remains unchanged in a $C_4$-symmetric square lattice: two successive flips
}
A particularly instructive multiband case is provided by a square-lattice TiO$_2$ photonic-crystal slab, for which the winding-flip rule is experimentally tested and shown for multiple successive inversions. The fabricated structure consists of a TiO$_2$ slab of period $a=400$~nm and thickness $h=100$~nm, patterned with square air holes of size $l$ etched to a depth of $50$~nm (etching fraction $=0.5$). Owing to the $C_4$ symmetry of the square lattice, the two bright modes remain exactly degenerate at $\Gamma$, while two dark modes host symmetry-protected BICs. The band ordering can therefore be classified as DDBB, DBBD, or BBDD depending on the ratio $l/a$, which serves as the inversion control parameter.

As $l/a$ is varied, the degenerate bright pair exchanges its ordering successively with the dark branches. In other words, each dark mode effectively undergoes two consecutive dark/bright inversions. According to the general rule established above, each individual inversion flips the winding carried by the corresponding band-edge BIC. Because the dark mode ``jumps'' across two bright partners in sequence, its winding flips twice and therefore returns to its initial value. The net result is that, although band inversion occurs, the winding number of a given dark branch remains unchanged between the DDBB and BBDD configurations.

This behavior is confirmed experimentally by angle-resolved reflectivity measurements and by the extracted far-field polarization textures of the dark bands. As shown in Fig.~\ref{fig:C4_square_lattice}, the measured textures are in excellent agreement with both the analytical predictions using an effective $4\times4$ non-Hermitian Hamiltonian~\cite{nguyen_generalized_2025} and with RCWA simulations. The intermediate DBBD configuration clearly reveals the two-step exchange process, while the initial and final configurations demonstrate that the overall winding of each dark branch is preserved after the two successive flips. This square-lattice example therefore provides a direct multiband verification of the consistency of the winding-flip rule: what matters is not simply that inversions occur, but how many bright partners are crossed by a given dark branch.

\subsection*{Multiple band inversions in a triangular lattice}

As a further test of the generality of the winding-flip rule, we consider a multiband triangular-lattice photonic-crystal slab and analyze its band inversions numerically using RCWA. The structure is analogous to the rectangular-lattice platform discussed in the main text and to the square-lattice case presented above: a TiO$_2$ slab of thickness $h=100$~nm on a glass substrate, etched to a depth of $50$~nm, with lattice period $a=400$~nm. The unit cell contains an elliptical air hole with principal axes $l_x$ and $l_y$ [Fig.~\ref{fig:triangular_lattice}(a)]. The reciprocal lattice and high-symmetry points are shown in Fig.~\ref{fig:triangular_lattice}(b).

Two symmetry classes can be distinguished. When $l_x=l_y$, the air hole is circular and the structure preserves full $C_6$ symmetry. In that case, as discussed in \cite{nguyen_generalized_2025}, the six lowest bands near $\Gamma$ consist of two degenerate bright modes of magnetic-dipolar character and four dark modes: two degenerate magnetic-quadrupolar modes, each hosting a first-order BIC, together with two singly degenerate modes, namely a monopolar first-order BIC and a hexapolar second-order BIC. When $l_x\neq l_y$, the $C_6$ symmetry is broken, but two $C_2$ mirror symmetries remain. The number of dark and bright bands is unchanged, but all degeneracies are lifted, in close analogy with the rectangular-lattice case of the main text.

Band inversions are induced by tuning the hole axes $l_x$ and $l_y$, which modify both the near-field hybridization and the radiative coupling of the six bands. The corresponding RCWA absorption spectra are shown in Fig.~\ref{fig:triangular_lattice}(c), together with the momentum-space polarization textures used to extract the winding numbers. In the $C_6$-symmetric case, the situation closely parallels the $C_4$-symmetric square lattice discussed in the previous section: because the two bright modes are degenerate, whenever a dark branch exchanges its ordering with the bright doublet it effectively undergoes a double jump across two bright partners. Each individual dark--bright inversion flips the winding of the corresponding band-edge BIC, but the two successive flips compensate each other. As a result, the winding numbers of the dark bands remain unchanged across all the $C_6$-symmetric configurations considered here.

By contrast, once the $C_6$ symmetry is broken and the bright-mode degeneracy is lifted, the individual inversions can be resolved one by one. In that regime, the winding number of a dark band flips sign after each dark--bright band inversion, exactly as expected from the general rule established in the main text. The triangular-lattice platform therefore provides a six-band numerical verification of the same principle: degenerate bright partners lead to successive compensated flips, whereas lifting the degeneracy restores the one-by-one sign reversal of the BIC winding at each inversion.

\subsection*{Coexistence of symmetry-protected and Friedrich--Wintgen BICs on a single branch}

\begin{figure}[hbt!]
\begin{center}
\includegraphics[width=0.8\linewidth]{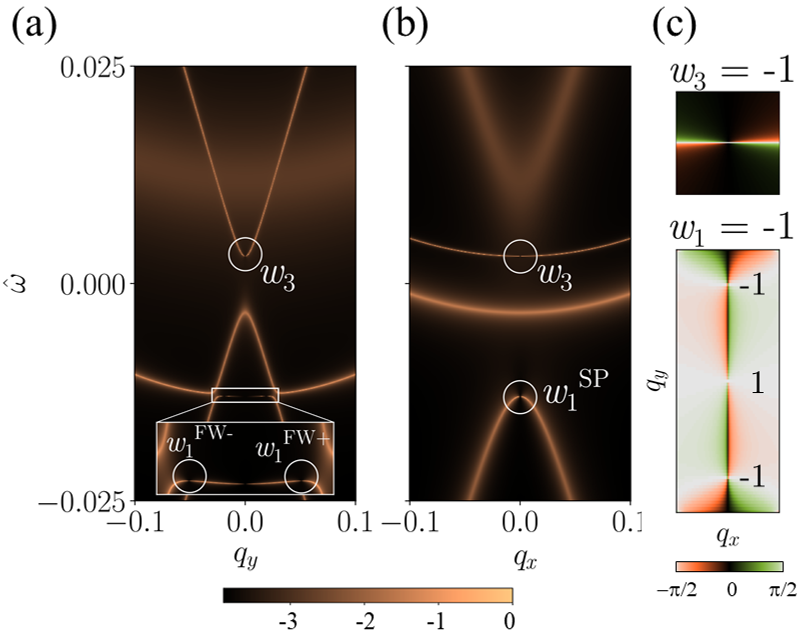}
\caption{\textbf{Coexistence of symmetry-protected and Friedrich--Wintgen BICs on the same branch.}
Simulated results for $l_x/a=0.3$, $l_y/a=0.856$, and $\beta=0$.
(a) Angle-resolved absorption spectrum along $\Gamma Y$.
The first band hosts one symmetry-protected (SP) BIC at $\Gamma$ and two off-$\Gamma$ Friedrich--Wintgen (FW) BICs (see inset).
(b) Absorption spectrum along $\Gamma X$.
(c) Polarization-orientation textures for the third band (top) and the first band (bottom).
The momentum ranges are $q_x\in[-0.01,0.01]$; for the third band $q_y\in[-0.1,0.1]$, and for the first band $q_y\in[-0.017,0.017]$.}
\label{fig:FWSP}
\end{center}
\end{figure}

The winding-flip rule established in the main text applies branch-by-branch and remains valid even when several BICs coexist on a single branch. Figure~\ref{fig:FWSP} illustrates this in the rectangular-lattice platform, for the parameters $l_x/a=0.3$, $l_y/a=0.856$, $\beta=0$. The first band hosts a symmetry-protected BIC at $\Gamma$ and two off-$\Gamma$ Friedrich--Wintgen BICs along $q_y$, all three carrying their own polarization vortex. Their algebraic sum gives a total branch-resolved winding $w_1=-1$, and the third band carries $w_3=-1$, consistent with the BDBD configuration in Fig.~\ref{fig:bandperm_2D}(b) of the main text. What flips under inversion is therefore not the winding of an individual singularity but the \emph{branch-resolved total} winding: a generalization of the two-band rule that accommodates multiple coexisting singularities on the same band.

\subsection*{Compact rule for multiband band inversions}

The rule established in the main text and verified above can be summarized as follows. Consider a multiband non-Hermitian band structure near $\Gamma$ in which dark (D) branches host symmetry-protected BICs and bright (B) branches are radiative. For each elementary band inversion that exchanges a dark branch $i$ with a non-degenerate bright branch $j$, the BIC winding on branch $i$ flips sign,
\begin{equation}
w_i \ \longrightarrow\ -\,w_i.
\label{eq:rule_single}
\end{equation}
When the bright partner is doubly degenerate (as in $C_4$- or $C_6$-symmetric lattices), the dark branch necessarily crosses both partners in succession, producing two compensating flips and leaving the winding unchanged,
\begin{equation}
w_i \ \longrightarrow\ -w_i \ \longrightarrow\ +w_i.
\label{eq:rule_double}
\end{equation}
What ultimately controls the net change of $w_i$ is therefore not whether band inversions occur, but how many bright partners a given dark branch crosses --- modulo the degeneracy multiplicity dictated by the lattice symmetry.


\onecolumngrid

\begin{center}
		\textbf{\large --- Supplementary Material ---\\ Band Inversion Flips the Winding of Bound States in the Continuum}\\
		\medskip
	\end{center}

\setcounter{equation}{0}
\setcounter{figure}{0}
\setcounter{table}{0}
\setcounter{page}{1}

\renewcommand{\theequation}{S\arabic{equation}}
\renewcommand{\thefigure}{S\arabic{figure}}
\renewcommand{\bibnumfmt}[1]{[S#1]}
\renewcommand{\vec}[1]{\boldsymbol{#1}}
\vspace{3cm}



\section{Derivation of the linearized far-field components near a collected BIC}

This Supplemental Note provides the detailed steps leading to Eqs.~(1)--(4) in the main text for the linearized collected far field
$E_{\parallel}$ and $E_{\perp}$ in the vicinity of a BIC, together with explicit expressions for the coefficients
$\mathbf g_{\perp}$ and $\mathbf g_{\parallel}$.

\subsection*{General projected radiation map and BIC condition}

We consider two nearby Bloch resonances (basis states) $|\Psi_1(\vec{k})\rangle$ and $|\Psi_2(\vec{k})\rangle$ forming an effective
two-band description near a band edge. The relevant eigenmode (say, the upper branch) is
\begin{equation}
|\Psi_{+}(\vec{k})\rangle=\alpha_1(\vec{k})|\Psi_1(\vec{k})\rangle+\alpha_2(\vec{k})|\Psi_2(\vec{k})\rangle,
\label{eq:S_eigvec}
\end{equation}
with complex coefficients $\alpha_m(\vec{k})$ defined up to an overall gauge.

Because the structure is open, each basis mode radiates into a set of open channels $c$ (diffraction orders, up/down ports, and polarizations).
Denote by $D_{cm}(\vec{k})$ the complex radiation amplitude from basis mode $m$ into channel $c$, and collect them into a channel vector
$\mathbf D_m(\vec{k})=(D_{cm}(\vec{k}))_{c\in{\rm open}}$.
In experiments and in polarization maps one typically analyzes a \emph{collected} field (e.g. a selected diffraction order within a numerical aperture),
which we represent by a linear projection $\mathcal P$ onto the collected subspace. The corresponding measured Jones vector is
\begin{equation}
\mathbf E(\vec{k}) \equiv \mathcal P\,\mathbf E_{\rm out}(\vec{k})
=\sum_{m=1,2}\alpha_m(\vec{k})\,\mathbf d_m(\vec{k}),
\qquad
\mathbf d_m(\vec{k})\equiv \mathcal P\,\mathbf D_m(\vec{k}).
\label{eq:S_measured_field}
\end{equation}
Equation~\eqref{eq:S_measured_field} is general: it includes multiple diffraction orders and polarizations and reduces to a single-channel description when
$\mathcal P$ selects a single output channel.

Defining the mixing ratio $r(\vec{k})\equiv\alpha_1(\vec{k})/\alpha_2(\vec{k})$, we rewrite Eq.~\eqref{eq:S_measured_field} as
\begin{equation}
\mathbf E(\vec{k})=\alpha_2(\vec{k})\Big[r(\vec{k})\,\mathbf d_1(\vec{k})+\mathbf d_2(\vec{k})\Big].
\label{eq:S_E_r_form}
\end{equation}
A collected BIC at $\vec{k}=\vec{k}^{\rm BIC}$ is defined by $\mathbf E(\vec{k}^{\rm BIC})=\mathbf 0$, hence
\begin{equation}
r^{\rm BIC}\mathbf d_1^{\rm BIC}+\mathbf d_2^{\rm BIC}=\mathbf 0,
\qquad
\mathbf d_2^{\rm BIC}=-r^{\rm BIC}\mathbf d_1^{\rm BIC},
\label{eq:S_BIC_collinear}
\end{equation}
where the superscript ``BIC'' denotes evaluation at $\vec{k}^{\rm BIC}$.

\subsection*{Linearization of the collected field around the BIC}

Let $\delta\vec{k}=\vec{k}-\vec{k}^{\rm BIC}$. Starting from
\begin{equation}
\mathbf E(\vec{k})=\alpha_2(\vec{k})\Big[r(\vec{k})\,\mathbf d_1(\vec{k})+\mathbf d_2(\vec{k})\Big],
\label{eq:S_E_r_form}
\end{equation}
we perform a first-order Taylor expansion around $\vec{k}^{\rm BIC}$. Introducing the shorthand
\[
\alpha_2(\vec{k}^{\rm BIC}+\delta\vec{k})=\alpha_2^{\rm BIC}+(\nabla_{\vec{k}}\alpha_2)_{\rm BIC}\cdot\delta\vec{k}+O(|\delta\vec{k}|^2),
\]
\[
r(\vec{k}^{\rm BIC}+\delta\vec{k})=r^{\rm BIC}+(\nabla_{\vec{k}}r)_{\rm BIC}\cdot\delta\vec{k}+O(|\delta\vec{k}|^2),
\]
\[
\mathbf d_m(\vec{k}^{\rm BIC}+\delta\vec{k})=\mathbf d_m^{\rm BIC}+(\nabla_{\vec{k}}\mathbf d_m)_{\rm BIC}\cdot\delta\vec{k}+O(|\delta\vec{k}|^2),
\]
and substituting into Eq.~\eqref{eq:S_E_r_form}, we obtain
\begin{align}
\mathbf E(\vec{k}^{\rm BIC}+\delta\vec{k})
&=
\Big[\alpha_2^{\rm BIC}+(\nabla_{\vec{k}}\alpha_2)_{\rm BIC}\cdot\delta\vec{k}\Big]
\Big\{
\Big[r^{\rm BIC}+(\nabla_{\vec{k}}r)_{\rm BIC}\cdot\delta\vec{k}\Big]
\Big[\mathbf d_1^{\rm BIC}+(\nabla_{\vec{k}}\mathbf d_1)_{\rm BIC}\cdot\delta\vec{k}\Big]
\nonumber\\
&\hspace{3.6cm}
+\Big[\mathbf d_2^{\rm BIC}+(\nabla_{\vec{k}}\mathbf d_2)_{\rm BIC}\cdot\delta\vec{k}\Big]
\Big\}
+O(|\delta\vec{k}|^2).
\label{eq:S_taylor_full}
’\end{align}
Keeping only the terms linear in $\delta\vec{k}$ and using the BIC condition
$r^{\rm BIC}\mathbf d_1^{\rm BIC}+\mathbf d_2^{\rm BIC}=\mathbf 0$, all contributions proportional to
$(\nabla_{\vec{k}}\alpha_2)_{\rm BIC}\cdot\delta\vec{k}$ drop out because they multiply the vanishing bracket
$[r^{\rm BIC}\mathbf d_1^{\rm BIC}+\mathbf d_2^{\rm BIC}]$. One therefore finds
\begin{equation}
\boxed{
\mathbf E(\vec{k}^{\rm BIC}+\delta\vec{k})
=
\alpha_2^{\rm BIC}\Big[
\big((\nabla_{\vec{k}}r)_{\rm BIC}\cdot\delta\vec{k}\big)\,\mathbf d_1^{\rm BIC}
+
\big(\nabla_{\vec{k}}[\,r^{\rm BIC}\mathbf d_1+\mathbf d_2\,]\big)_{\rm BIC}\cdot\delta\vec{k}
\Big]
+O(|\delta\vec{k}|^2).
}
\label{eq:S_linear_compact}
\end{equation}
The quantity 
$\big(\nabla_{\vec{k}}[\,r^{\rm BIC}\mathbf d_1+\mathbf d_2\,]\big)_{\rm BIC}$
denotes the Jacobian matrix of the polarization vector
$\mathbf F(\vec k)=r^{\rm BIC}\mathbf d_1(\vec k)+\mathbf d_2(\vec k)$
with respect to the in-plane momentum $\vec k=(k_x,k_y)$.
Since $\mathbf F(\vec k)$ is a two-component Jones vector,
its gradient with respect to $\vec k$ is a $2\times2$ matrix,
\[
\nabla_{\vec k}\mathbf F
=
\begin{pmatrix}
\partial_{k_x}F_x & \partial_{k_y}F_x \\
\partial_{k_x}F_y & \partial_{k_y}F_y
\end{pmatrix},
\]
evaluated at $\vec k^{\rm BIC}$. The contraction with
$\delta\vec k=(\delta k_x,\delta k_y)^T$
corresponds to matrix–vector multiplication,
\[
\big(\nabla_{\vec{k}}\mathbf F\big)_{\rm BIC}\cdot\delta\vec k
=
\begin{pmatrix}
\partial_{k_x}F_x\,\delta k_x + \partial_{k_y}F_x\,\delta k_y \\
\partial_{k_x}F_y\,\delta k_x + \partial_{k_y}F_y\,\delta k_y
\end{pmatrix}_{\rm BIC},
\]
which yields a polarization vector describing the linear change of the collected radiation field induced by a small displacement in momentum space. This term therefore captures the smooth $k$-dependence of the projected radiation basis $\mathbf d_{1,2}(\vec k)$ and is independent of the eigenvector mixing ratio $r(\vec k)$.

\subsection*{Projection onto a fixed local polarization basis: $E_{\parallel}$ and $E_{\perp}$}

We now choose an orthonormal polarization basis $\{\hat{\mathbf e}_{\parallel},\hat{\mathbf e}_{\perp}\}$ \emph{fixed at the BIC} such that
\begin{equation}
\hat{\mathbf e}_{\parallel}\parallel \mathbf d_1^{\rm BIC},
\qquad
\hat{\mathbf e}_{\perp}\perp \hat{\mathbf e}_{\parallel}.
\label{eq:S_pol_basis}
\end{equation}
In this basis, the collected field is decomposed as
$\mathbf E(\vec{k})=E_{\parallel}(\vec{k})\,\hat{\mathbf e}_{\parallel}+E_{\perp}(\vec{k})\,\hat{\mathbf e}_{\perp}$ with scalar components
\begin{equation}
E_{\parallel,\perp}(\vec{k})\equiv \hat{\mathbf e}_{\parallel,\perp}\cdot\mathbf E(\vec{k}),
\qquad
d_{m,\parallel/\perp}(\vec{k})\equiv \hat{\mathbf e}_{\parallel/\perp}\cdot\mathbf d_m(\vec{k}).
\label{eq:S_comp_defs}
\end{equation}
By construction, $d_{1,\perp}^{\rm BIC}=0$. Moreover, the BIC condition
$r^{\rm BIC}\mathbf d_1^{\rm BIC}+\mathbf d_2^{\rm BIC}=0$ implies that $\mathbf d_2^{\rm BIC}$ is collinear with $\mathbf d_1^{\rm BIC}$,
hence also $d_{2,\perp}^{\rm BIC}=0$.

Starting from the linearized field in Eq.~\eqref{eq:S_linear_compact} and projecting onto $\hat{\mathbf e}_{\perp}$ gives
\begin{align}
E_{\perp}(\vec{k}^{\rm BIC}+\delta\vec{k})
&=\hat{\mathbf e}_{\perp}\cdot\mathbf E(\vec{k}^{\rm BIC}+\delta\vec{k})
\nonumber\\
&=\alpha_2^{\rm BIC}\Big[
\big((\nabla_{\vec{k}}r)_{\rm BIC}\cdot\delta\vec{k}\big)\underbrace{(\hat{\mathbf e}_{\perp}\cdot\mathbf d_1^{\rm BIC})}_{=\,d_{1,\perp}^{\rm BIC}=0}
+
\hat{\mathbf e}_{\perp}\cdot
\Big(\big(\nabla_{\vec{k}}[\,r^{\rm BIC}\mathbf d_1+\mathbf d_2\,]\big)_{\rm BIC}\cdot\delta\vec{k}\Big)
\Big]
+O(|\delta\vec{k}|^2).
\end{align}
The $\nabla_{\vec k}r$ term vanishes identically because $d_{1,\perp}^{\rm BIC}=0$. Using the component notation
$d_{m,\perp}(\vec{k})=\hat{\mathbf e}_{\perp}\cdot\mathbf d_m(\vec{k})$, we obtain the generic linear form
\begin{equation}
\boxed{
E_{\perp}(\vec{k}^{\rm BIC}+\delta\vec{k})
=
\mathbf g_{\perp}\cdot\delta\vec{k}
+O(|\delta\vec{k}|^2),
\qquad
\mathbf g_{\perp}
=
\alpha_2^{\rm BIC}
\Big(\nabla_{\vec{k}}\big[\,r^{\rm BIC}d_{1,\perp}(\vec{k})+d_{2,\perp}(\vec{k})\,\big]\Big)_{\rm BIC}.
}
\label{eq:S_Eperp_gperp}
\end{equation}
For an isolated first-order vortex (generic $|w_{\rm BIC}|=1$), $\mathbf g_{\perp}\neq\mathbf 0$; otherwise higher-order terms control the texture.

Similarly, projecting Eq.~\eqref{eq:S_linear_compact} onto $\hat{\mathbf e}_{\parallel}$ yields
\begin{align}
E_{\parallel}(\vec{k}^{\rm BIC}+\delta\vec{k})
&=\hat{\mathbf e}_{\parallel}\cdot\mathbf E(\vec{k}^{\rm BIC}+\delta\vec{k})
\nonumber\\
&=\alpha_2^{\rm BIC}\Big[
\big((\nabla_{\vec{k}}r)_{\rm BIC}\cdot\delta\vec{k}\big)\,(\hat{\mathbf e}_{\parallel}\cdot\mathbf d_1^{\rm BIC})
+
\hat{\mathbf e}_{\parallel}\cdot
\Big(\big(\nabla_{\vec{k}}[\,r^{\rm BIC}\mathbf d_1+\mathbf d_2\,]\big)_{\rm BIC}\cdot\delta\vec{k}\Big)
\Big]
+O(|\delta\vec{k}|^2).
\end{align}
Defining $d_{m,\parallel}(\vec{k})=\hat{\mathbf e}_{\parallel}\cdot\mathbf d_m(\vec{k})$ and
\begin{equation}
A\equiv \alpha_2^{\rm BIC}d_{1,\parallel}^{\rm BIC}\neq 0,
\label{eq:S_A_def}
\end{equation}
we obtain
\begin{equation}
\boxed{
E_{\parallel}(\vec{k}^{\rm BIC}+\delta\vec{k})
=
A\big((\nabla_{\vec{k}}r)_{\rm BIC}\cdot\delta\vec{k}\big)
+\mathbf g_{\parallel}\cdot\delta\vec{k}
+O(|\delta\vec{k}|^2),
\qquad
\mathbf g_{\parallel}
=
\alpha_2^{\rm BIC}
\Big(\nabla_{\vec{k}}\big[\,r^{\rm BIC}d_{1,\parallel}(\vec{k})+d_{2,\parallel}(\vec{k})\,\big]\Big)_{\rm BIC}.
}
\label{eq:S_Epar_gpar}
\end{equation}
The first term in Eq.~\eqref{eq:S_Epar_gpar} is the $\nabla_{\vec k}r$ contribution arising from the $\vec k$-dependence of the mixing ratio $r(\vec{k})$, whereas $\mathbf g_{\parallel}$ and $\mathbf g_{\perp}$ originate from the smooth $\vec k$-dependence of the projected radiation vectors $\mathbf d_{1,2}(\vec{k})$.

\subsection*{Basis choice at the BIC and decomposition $r=|r|e^{i\beta}$}

Write $r(\vec{k})=|r(\vec{k})|e^{i\beta(\vec{k})}$. The BIC cancellation condition in the collected channel,
$r^{\rm BIC}\mathbf d_1^{\rm BIC}+\mathbf d_2^{\rm BIC}=\mathbf 0$, pins the relative phase between $\mathbf d_1^{\rm BIC}$ and $\mathbf d_2^{\rm BIC}$.
One may choose a the gauge in which both  $\mathbf d_1^{\rm BIC}$ and $\mathbf d_2^{\rm BIC}$ are real,  such that
\begin{equation}
\beta(\vec{k}^{\rm BIC})=0\ \ \text{or}\ \ \pi,
\qquad\text{equivalently}\qquad r^{\rm BIC}\in\mathbb R.
\label{eq:S_beta_gauge}
\end{equation}
Then the $\nabla_{\vec k}r$ term in Eq.~\eqref{eq:S_Epar_gpar} can be written explicitly as
\begin{equation}
(\nabla_{\vec{k}}r)_{\rm BIC}
=
e^{i\beta^{\rm BIC}}(\nabla_{\vec{k}}|r|)_{\rm BIC}
+i|r^{\rm BIC}|e^{i\beta^{\rm BIC}}(\nabla_{\vec{k}}\beta)_{\rm BIC},
\label{eq:S_grad_r_split}
\end{equation}
so that
\begin{equation}
\boxed{
E_{\parallel}(\vec{k}^{\rm BIC}+\delta\vec{k})
=
A'\,(\nabla_{\vec{k}}|r|)_{\rm BIC}\cdot\delta\vec{k}
+iA'|r^{\rm BIC}|\,(\nabla_{\vec{k}}\beta)_{\rm BIC}\cdot\delta\vec{k}
+\mathbf g_{\parallel}\cdot\delta\vec{k}
+O(|\delta\vec{k}|^2),
}
\label{eq:S_Epar_final}
\end{equation}
with $A'\equiv A\cos\beta^{\rm BIC}=\pm A$. This is the form used in the main text.

\vspace{1ex}
\noindent\textbf{Remark.} The derivation above assumes (i) an effective two-band description near the relevant band edge,
(ii) a collected BIC defined by $\mathbf E(\vec{k}^{\rm BIC})=\mathbf 0$, and (iii) an isolated first-order BIC vortex (generic $|w_{\rm BIC}|=1$),
so that the linear terms in Eqs.~\eqref{eq:S_Eperp_gperp}--\eqref{eq:S_Epar_gpar} control the local polarization texture.

\subsection*{Sign flip of \texorpdfstring{$\nabla_{\vec k}|r|$}{grad |r|} across a non-Hermitian band inversion}

The body's argument [Eq.~(\ref{eq:inversion_signature_main}) of the main text] hinges on the inversion-induced sign reversal of $\nabla_{\vec k}|r|^{\rm BIC}$. We show here how this sign flip arises in the open, non-Hermitian setting where the radiative loss remains finite, and we derive the regularized profile of $\nabla_{\vec k}|r|^{\rm BIC}$ across the inversion.

Writing $r(\vec k)=|r(\vec k)|\,e^{i\beta(\vec k)}$, the gradient decomposes as $\partial_{\vec k}r = e^{i\beta}\big[\nabla_{\vec k}|r| + i\,|r|\,\nabla_{\vec k}\beta\big]$. In the gauge of the previous subsection, $\beta^{\rm BIC}=0$ or $\pi$, so $e^{i\beta^{\rm BIC}}=\pm 1$ and the modulus gradient at the BIC is the real part of $\partial_{\vec k}r$:
\begin{equation}
\nabla_{\vec k}|r|^{\rm BIC} \;=\; \pm\,\Re(\partial_{\vec k}r)^{\rm BIC},\qquad
|r^{\rm BIC}|\,\nabla_{\vec k}\beta^{\rm BIC} \;=\; \pm\,\Im(\partial_{\vec k}r)^{\rm BIC}.
\label{eq:S_nabla_r_real}
\end{equation}
The modulus gradient and the phase gradient are therefore the real and imaginary channels of the same complex slope.

A standard $k\!\cdot\!p$ argument on the two-band non-Hermitian Hamiltonian gives the leading slope of the mixing ratio near a band-edge BIC,
\begin{equation}
\partial_{\vec k}r \;\propto\; \frac{1}{\Omega_+ - \Omega_-} \;=\; \frac{1}{\Delta_{\rm Re} + i\,\Gamma_{\rm rad}},
\label{eq:S_kp_slope}
\end{equation}
with $\Delta_{\rm Re}$ and $\Gamma_{\rm rad}$ the real and imaginary parts of the complex gap between the two non-Hermitian branches; for a band-edge BIC, $\Gamma_{\rm rad}$ reduces to the radiative loss rate of the bright partner. Splitting Eq.~\eqref{eq:S_kp_slope} into real and imaginary parts:
\begin{equation}
\Re(\partial_{\vec k}r) \;\propto\; \frac{\Delta_{\rm Re}}{\Delta_{\rm Re}^2+\Gamma_{\rm rad}^2},\qquad
\Im(\partial_{\vec k}r) \;\propto\; -\frac{\Gamma_{\rm rad}}{\Delta_{\rm Re}^2+\Gamma_{\rm rad}^2}.
\label{eq:S_kp_split}
\end{equation}

Two observations are immediate. The real part is an \emph{odd} function of $\Delta_{\rm Re}$ and therefore changes sign through the inversion ($\Delta_{\rm Re}\to-\Delta_{\rm Re}$), regardless of how large $\Gamma_{\rm rad}$ is: the radiative loss regularizes the magnitude but cannot affect the parity. The imaginary part is an \emph{even} function of $\Delta_{\rm Re}$ and remains finite throughout. Combining with Eq.~\eqref{eq:S_nabla_r_real},
\begin{equation}
\nabla_{\vec k}|r|^{\rm BIC} \;\propto\; \frac{\Delta_{\rm Re}}{\Delta_{\rm Re}^2+\Gamma_{\rm rad}^2},
\label{eq:S_nabla_r_lorentzderiv}
\end{equation}
which has a Lorentzian-derivative profile in $\Delta_{\rm Re}$: it peaks at $\Delta_{\rm Re}=\pm\,\Gamma_{\rm rad}$ with extremal magnitude $|\nabla_{\vec k}|r||_{\max}\sim 1/(2\Gamma_{\rm rad})$, vanishes smoothly at the inversion point $\Delta_{\rm Re}=0$, and reduces to the Hermitian $1/\Delta_{\rm Re}$ scaling for $|\Delta_{\rm Re}|\gg\Gamma_{\rm rad}$.

The inversion mechanism in the body therefore proceeds as follows: as the structural parameter is tuned, $\Delta_{\rm Re}$ passes through zero, $\nabla_{\vec k}|r|^{\rm BIC}$ continuously decreases from $+1/(2\Gamma_{\rm rad})$ through zero (at the inversion) on to $-1/(2\Gamma_{\rm rad})$. The zero crossing at $\Delta_{\rm Re}=0$ makes the linearized map $\delta\vec k\to\mathbf E$ rank-deficient at the inversion (the texture-collapse seen in the central panel of Fig.~1(b) of the main text), and the sign change before and after is the winding flip.

\subsection*{Vanishing of \texorpdfstring{$\mathbf g_\parallel$}{g\_parallel} for smooth radiation channels}

The body's analysis assumes that the radiation-channel amplitudes vary slowly enough near the BIC that $\mathbf g_\parallel\approx 0$ while $\mathbf g_\perp\neq 0$. We make this assumption precise here and verify it for the platforms studied.

\paragraph*{Geometric exact result.} If the radiation-channel magnitudes are stationary at the BIC, $\nabla_{\vec k}|\mathbf d_i|^{\rm BIC}=0$ for $i=1,2$, then $\mathbf g_\parallel=0$ exactly. The proof is geometric: from $|\mathbf d_i|^2=\mathbf d_i\cdot\mathbf d_i$,
\begin{equation}
\mathbf d_i^{\rm BIC}\cdot\nabla_{\vec k}\mathbf d_i^{\rm BIC}\;=\;\tfrac{1}{2}\nabla_{\vec k}|\mathbf d_i|^2\big|_{\rm BIC}\;=\;0.
\label{eq:S_d_smooth}
\end{equation}
By construction $\hat{\mathbf e}_\parallel\parallel\mathbf d_1^{\rm BIC}$, and the BIC condition $\mathbf d_2^{\rm BIC}=-r^{\rm BIC}\mathbf d_1^{\rm BIC}$ implies $\hat{\mathbf e}_\parallel\parallel\mathbf d_2^{\rm BIC}$ as well. Hence
\begin{equation}
\nabla_{\vec k}d_{i,\parallel}^{\rm BIC}\;=\;\hat{\mathbf e}_\parallel\cdot\nabla_{\vec k}\mathbf d_i^{\rm BIC}\;=\;\frac{\mathbf d_i^{\rm BIC}\cdot\nabla_{\vec k}\mathbf d_i^{\rm BIC}}{|\mathbf d_i^{\rm BIC}|}\;=\;0,
\label{eq:S_dpar_zero}
\end{equation}
and substitution into Eq.~\eqref{eq:S_Epar_gpar} gives $\mathbf g_\parallel=0$ identically. Geometrically: with $|\mathbf d_i|$ constant, the local variation of $\mathbf d_i$ is purely \emph{transverse} to $\mathbf d_i$ itself --- and hence transverse to $\hat{\mathbf e}_\parallel$. It contributes only to $\mathbf g_\perp$, never to $\mathbf g_\parallel$.

\paragraph*{Parametric argument when \texorpdfstring{$\nabla_{\vec k}|\mathbf d_i|^{\rm BIC}\neq 0$}{the channel magnitudes vary}.} For channels whose magnitude does have non-trivial $\vec k$-dependence, $\mathbf g_\parallel\neq 0$ in general. The winding flip nevertheless survives provided the inversion-driven contribution to $E_\parallel$ dominates the static $\mathbf g_\parallel$ contribution. Comparing the two scales:
\begin{itemize}\setlength{\itemsep}{0pt}
\item The inversion-driven slope $|\partial_{\vec k}r|^{\rm BIC}$ is bounded above by $v_g/(2\Gamma_{\rm rad})$ at $|\Delta_{\rm Re}|=\Gamma_{\rm rad}$ [Eq.~\eqref{eq:S_nabla_r_lorentzderiv}], where $v_g$ is the band group velocity.
\item The smooth radiation-channel gradient is bounded by the Brillouin-zone-scale variation, $|\nabla_{\vec k}\mathbf d_i^{\rm BIC}|\sim|\mathbf d_i^{\rm BIC}|\cdot a$, where $a$ is the lattice constant.
\end{itemize}
Their ratio defines the relevant figure of merit:
\begin{equation}
\frac{|A'\,\nabla_{\vec k}|r||}{|\mathbf g_\parallel|}\;\sim\;\frac{v_g}{\Gamma_{\rm rad}\,a}\;\sim\;\frac{Q}{n_g},
\label{eq:S_parametric_ratio}
\end{equation}
where $Q=\omega/(2\Gamma_{\rm rad})$ is the quality factor of the bright partner and $n_g=c/v_g$ is the band group index. For typical photonic-crystal-slab BICs, $Q\sim 10$--$10^3$ and $n_g\sim 1$--$10$, so $Q/n_g\gg 1$ --- the inversion-driven contribution dominates parametrically. The rank-deficient critical point at $\Delta_{\rm Re}=0$ is preserved up to corrections of relative order $n_g/Q$, and the winding flip survives.

\paragraph*{Verification in the systems studied.} For the 1D grating model worked out below, the unit-magnitude TE polarization vectors $\vec u_{\pm 1}(\vec q)$ have $|\vec u_{\pm 1}|=1$ identically, hence $\nabla_{\vec k}|\mathbf d_i|^{\rm BIC}=0$ exactly and $\mathbf g_\parallel=0$ exactly. This is consistent with the closed-form far-field of Eq.~\eqref{eq:E_reduced_revised}, which contains no $\mathbf g_\parallel\!\cdot\delta\vec k$ piece. For the multiband rectangular and triangular slabs of the main text, the achiral $C_2$ (resp.\ $C_6$) symmetry constrains $\nabla_{\vec k}|\mathbf d_i|^{\rm BIC}$ to be small, and the parametric ratio $Q/n_g$ is comfortably large in the simulated configurations. The flip is therefore robust both geometrically and parametrically in all the platforms considered.

\section{Winding number from Stokes parameters and a local Jacobian}

Near an isolated polarization singularity (BIC) at $\vec{k}_0$, the in-plane far-field Jones vector
$\mathbf E(\vec{k})=(E_x,E_y)$ vanishes, $\mathbf E(\vec{k}_0)=0$, and can be expanded as
$E_x=\mathbf a\!\cdot\!\delta\vec{k}$ and $E_y=\mathbf b\!\cdot\!\delta\vec{k}$ to leading order,
with $\delta\vec{k}=\vec{k}-\vec{k}_0$ and $\mathbf a,\mathbf b\in\mathbb C^2$.

To define a gauge-invariant polarization orientation, we use the Stokes parameters
\begin{equation}
S_1=|E_x|^2-|E_y|^2,\qquad
S_2=2\,\Re(E_xE_y^*),
\label{eq:stokes12}
\end{equation}
and introduce the two-component field $\mathbf s(\vec{k})=(S_1(\vec{k}),S_2(\vec{k}))$.
The integer vortex charge associated with the polarization texture is defined as the winding of $\mathbf s$ around the singularity,
\begin{equation}
w_{\rm BIC}
=\frac{1}{2\pi}\oint_{\mathcal C}\nabla_{\vec{k}}\arg\!\big(S_1+iS_2\big)\cdot d\vec{k}
\label{eq:w_stokes_def}
\end{equation}
where $\mathcal C$ is a small loop enclosing $\vec{k}_0$.

Using the linearized Jones field, the leading Stokes variations are quadratic in $\delta\vec{k}$:
\begin{align}
S_1(\delta\vec{k})&=
|\mathbf a\!\cdot\!\delta\vec{k}|^2-|\mathbf b\!\cdot\!\delta\vec{k}|^2, \label{eq:S1_quad}\\
S_2(\delta\vec{k})&=
2\,\Re\!\Big[(\mathbf a\!\cdot\!\delta\vec{k})\,(\mathbf b\!\cdot\!\delta\vec{k})^*\Big]. \label{eq:S2_quad}
\end{align}
The winding can be obtained locally without evaluating any line integral by considering the Jacobian of the map
$\vec{k}\mapsto(S_1,S_2)$ along $\mathcal C$. Specifically, define
\begin{equation}
J_s(\vec{k})=
\begin{pmatrix}
\partial_{k_x}S_1 & \partial_{k_y}S_1\\
\partial_{k_x}S_2 & \partial_{k_y}S_2
\end{pmatrix}.
\label{eq:Js_def}
\end{equation}
For a first-order (isolated) polarization singularity, the sign of $\det J_s$ is constant on a sufficiently small circle
$\mathcal C$ around $\vec{k}_0$ and fixes the handedness with which $\mathbf s(\mathcal C)$ winds around the origin. Consequently,
\begin{equation}
w_{\rm BIC}=\mathrm{sgn}\!\left[\det J_s\right]_{\mathcal C}.
\label{eq:w_from_Js}
\end{equation}

If $\det J_s$ vanishes on $\mathcal C$, the image $\mathbf s(\mathcal C)$ collapses and the singularity is no longer isolated (or becomes higher order),
so the winding cannot be assigned within the first-order description. This corresponds to the inversion point where the vortex temporarily loses its rotational character before re-emerging with opposite handedness.

\begin{figure}[!ht]
\begin{center}
\includegraphics[width=0.8 \textwidth]{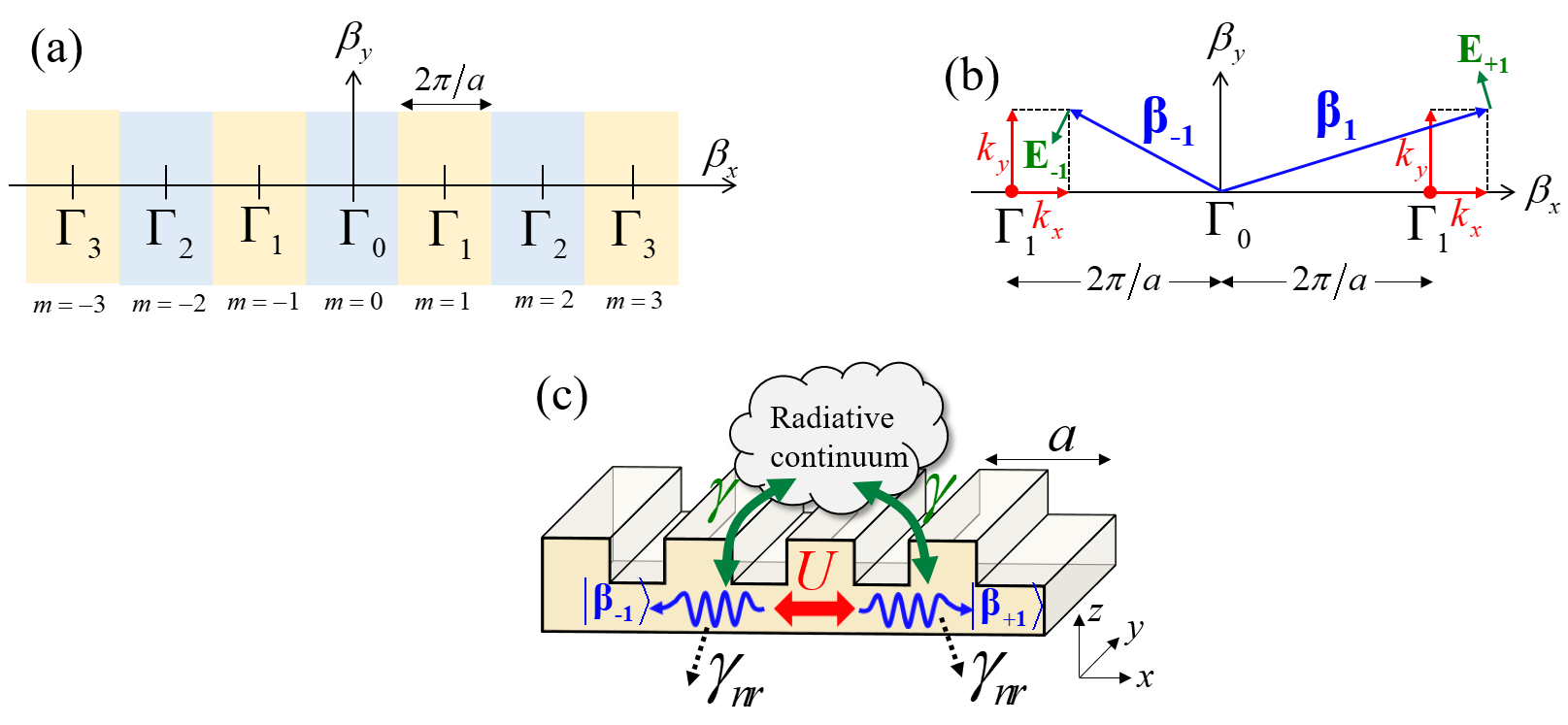}	
\caption{\textbf{Coupled-mode theory near the lowest $\Gamma$ point of a 1D grating.}
(a) Schematic of momentum space near the first $\Gamma$ point for a lattice of period $a$ along $x$ and invariant along $y$. The neighboring folded zones are labeled by the Brillouin-zone index $m\in\mathbb{Z}$. 
(b) A Bloch wavevector $(k_x,k_y)$ near $\Gamma$ is associated with two guided-mode wavevectors, $\vec{\beta}_{\pm1}=\left(k_x\pm2\pi/a\right)\vec{u}_x+k_y\vec{u}_y$, corresponding to the two counter-propagating TE guided waves with electric fields $\mathbf{E}_{\pm1}$. 
(c) Sketch of the coupling mechanisms entering the effective non-Hermitian Hamiltonian: coherent diffractive coupling of strength $U$ between $\ket{\vec{\beta}_{-1}}$ and $\ket{\vec{\beta}_{+1}}$, radiative coupling to the common continuum with decay rate $\gamma$, and possible non-radiative losses with rate $\gamma_{nr}$.}
\label{fig:CMT}
\end{center}
\end{figure}

\section{Analytical model for the photonic grating near $\Gamma$}

In this section, we derive in detail the effective non-Hermitian Hamiltonian describing the leaky Bloch modes of a 1D photonic grating of period $a$ along the $x$ axis, invariant along $y$, and preserving the lateral mirror symmetry $x\rightarrow -x$. This model is a particular case of the general non-Hermitian Hamiltonian for guided resonances in photonic crystal slabs introduced in ~\cite{nguyen_generalized_2025}, but we rederive it here explicitly in the simple grating geometry in order to make the origin of the BIC winding flip fully transparent. Note that here we adopt the time dependence $e^{-i\omega t}$. With this convention, losses correspond to a negative imaginary part of the complex eigenfrequency, so radiative and non-radiative decay rates appear in the Hamiltonian through terms proportional to $-i\gamma$ and $-i\gamma_{nr}$.

\subsection{Guided-mode basis near the first $\Gamma$ point}

In a perturbative description, the Bloch resonances of the grating are expanded in the basis of in-plane guided modes $\ket{\vec{\beta}}$ of the corresponding effective homogeneous slab. Close to the first $\Gamma$ point, the relevant states are obtained by folding guided modes from the neighboring Brillouin zones $m=\pm1$ back to the origin. Writing the Bloch wavevector as $\vec{k}=k_x\vec{u}_x+k_y\vec{u}_y$ with $|k_x|,|k_y|\ll 2\pi/a$, the corresponding in-plane propagation vectors are $\vec{\beta}_{m}(\vec{k})=\vec{k}+\frac{2\pi m}{a}\vec{u}_x$. Restricting to the vicinity of the lowest $\Gamma$ point, we keep only the pair
\begin{equation}
\vec{\beta}_{\pm1}(\vec{k})=\vec{k}\pm\frac{2\pi}{a}\vec{u}_x.
\label{eq:beta_pm1_revised}
\end{equation}
These two counter-propagating guided waves form the natural basis for the effective two-band description, as illustrated in Fig.~\ref{fig:CMT}(a,b).

\subsection{Dispersion of the folded guided modes}

We assume that, in the spectral range of interest, the dispersion of the guided mode of the effective slab is approximately linear in the modulus of the in-plane propagation vector, with slope determined by an effective group index $n_g$. Expanding around the folded wavevector magnitude $2\pi/a$, one obtains
\begin{align}
\omega_{\pm1}(\vec{k})
&=\omega_{\Gamma_1}+\frac{c}{n_g}\left(|\vec{\beta}_{\pm1}(\vec{k})|-\frac{2\pi}{a}\right)\nonumber\\
&\approx \omega_{\Gamma_1}\pm\frac{c}{n_g}k_x+\frac{ca}{4\pi n_g}k_y^2,
\label{eq:dispersion_dimensional}
\end{align}
where $\omega_{\Gamma_1}$ denotes the frequency of the folded guided mode at the first $\Gamma$ point.

We now introduce the dimensionless reduced momentum
\begin{equation}
\vec{q}=\frac{a}{2\pi}\vec{k},
\end{equation}
and the dimensionless frequency detuning
\begin{equation}
\hat{\omega}_{\pm1}(\vec{q})=\frac{a}{2\pi c}\left[\omega_{\pm1}(\vec{k})-\omega_{\Gamma_1}\right].
\end{equation}
Equation~\eqref{eq:dispersion_dimensional} then becomes
\begin{equation}
\hat{\omega}_{\pm1}(\vec{q})\approx \pm\frac{q_x}{n_g}+\frac{q_y^2}{2n_g}.
\label{eq:omega_pm_reduced}
\end{equation}
Thus, near $\Gamma$, the two folded guided modes are split linearly along $q_x$ and remain degenerate to first order along $q_y$.

\subsection{Polarization vectors of the guided modes}

We restrict the analysis to TE-polarized guided modes, for which the electric field is in the plane of the slab and orthogonal to the propagation vector. In the spectral range of interest around the first $\Gamma$-point band edge considered here, the TM-polarized guided modes of the effective slab lie far in frequency from the dark/bright TE pair, so even at finite $k_y$ --- where the lattice symmetry would generically allow TE--TM coupling --- no TM mode is present in the vicinity of the gap to couple to. The two-band TE description used below is therefore self-consistent throughout the spectral and momentum range of interest. The polarization vectors associated with $\vec{\beta}_{\pm1}$ can be written as
\begin{equation}
\vec{u}_{\pm1}(\vec{q})=\cos\theta_{\pm1}\,\vec{u}_x+\sin\theta_{\pm1}\,\vec{u}_y,
\label{eq:u_pm_revised}
\end{equation}
with $\vec{u}_{\pm1}\cdot\vec{\beta}_{\pm1}=0$. This gives
\begin{equation}
\left(\cos\theta_{\pm1},\sin\theta_{\pm1}\right)
=
\frac{\left(-q_y,q_x\pm1\right)}{\sqrt{(q_x\pm1)^2+q_y^2}}.
\label{eq:theta_pm_revised}
\end{equation}

Near $\Gamma$, these expressions reduce to
\begin{equation}
\cos\theta_{\pm1}=-q_y+O(q^3),
\qquad
\sin\theta_{+1}=1-\frac{q_y^2}{2}+O(q^3),
\qquad
\sin\theta_{-1}=-1+\frac{q_y^2}{2}+O(q^3).
\label{eq:smallq_polarizations}
\end{equation}
Hence, at normal incidence the two guided modes are almost oppositely polarized along $y$, while both acquire the same small $x$ component proportional to $-q_y$ away from $\Gamma$.

\subsection{Effective non-Hermitian Hamiltonian}

In the basis $\Psi=\left(\ket{\vec{\beta}_{+1}},\ket{\vec{\beta}_{-1}}\right)$, the two folded guided modes are coupled by two distinct mechanisms. First, the periodic corrugation coherently couples the two counter-propagating waves. This produces a Hermitian diffractive coupling of strength $U$. Second, both guided modes couple to the same radiative continuum. This gives rise not only to diagonal radiative losses, but also to an anti-Hermitian off-diagonal coupling mediated by the shared radiation channels. If the structure contains absorptive materials, additional non-radiative losses are included through a rate $\gamma_{nr}$.

The resulting effective Hamiltonian reads
\begin{equation}
H(\vec{q})=
\begin{pmatrix}
\hat{\omega}_{+1} & U\\
U & \hat{\omega}_{-1}
\end{pmatrix}
-i
\begin{pmatrix}
\gamma+\gamma_{nr} & \gamma\cos\alpha\\
\gamma\cos\alpha & \gamma+\gamma_{nr}
\end{pmatrix},
\label{eq:H_full_revised}
\end{equation}
where $\gamma$ is the radiative decay rate of each uncoupled guided mode and
\begin{equation}
\cos\alpha(\vec{q})=\vec{u}_{-1}\cdot\vec{u}_{+1}
\end{equation}
is the overlap of their TE polarization vectors. This factor appears because only identical polarization components interfere in the radiative continuum.

Using Eq.~\eqref{eq:theta_pm_revised}, one obtains
\begin{equation}
\cos\alpha(\vec{q})
=
\frac{q_x^2+q_y^2-1}
{\sqrt{\big[(q_x-1)^2+q_y^2\big]\big[(q_x+1)^2+q_y^2\big]}}.
\label{eq:cosalpha_exact_revised}
\end{equation}
Near $\Gamma$, this reduces to
\begin{equation}
\cos\alpha(\vec{q})=-1+2q_y^2+O(q^4).
\label{eq:cosalpha_smallq_revised}
\end{equation}
Thus, to leading order, one may use $\cos\alpha\simeq -1$, so that
\begin{equation}
U-i\gamma\cos\alpha \simeq U+i\gamma.
\label{eq:d1_leading}
\end{equation}

It is convenient to rewrite Eq.~\eqref{eq:H_full_revised} in Pauli-matrix form,
\begin{equation}
H=d_0\mathbb{I}_2+d_1\sigma_1+d_3\sigma_3,
\label{eq:H_pauli_revised}
\end{equation}
with
\begin{align}
d_0&=\frac{q_y^2}{2n_g}-i(\gamma+\gamma_{nr}),\\
d_1&=U-i\gamma\cos\alpha,\\
d_3&=\frac{q_x}{n_g}.
\end{align}
This form separates the common frequency shift $d_0$, the coupling term $d_1$, and the odd-in-$q_x$ detuning $d_3$.

\subsection{Eigenvalues and band inversion}

The complex eigenvalues are
\begin{equation}
\Omega_\pm(\vec{q})
=
d_0\pm\sqrt{d_1^2+d_3^2}
=
\frac{q_y^2}{2n_g}-i(\gamma+\gamma_{nr})
\pm
\sqrt{
\left(\frac{q_x}{n_g}\right)^2+\left(U-i\gamma\cos\alpha\right)^2
}.
\label{eq:Omega_pm_revised}
\end{equation}

At $\Gamma$, where $q_x=q_y=0$ and $\cos\alpha=-1$, this becomes
\begin{equation}
\omega_\pm(0)=\pm U -i(\gamma\mp \gamma+\gamma_{nr}).
\label{eq:Omega_Gamma}
\end{equation}
The real parts are split by $\pm U$, so the sign of $U$ determines the ordering of the two branches. Tuning $U$ through zero therefore realizes a band inversion.

The imaginary parts are $\mathrm{Im}\,\omega_\pm(0)=-(\gamma\mp \gamma+\gamma_{nr})$, so the (+) branch has total loss $-\gamma_{nr}$ and the (-) branch has loss $-(2\gamma+\gamma_{nr})$. In the absence of non-radiative loss, the former has vanishing imaginary part and is therefore a symmetry-protected BIC.

\subsection{Eigenvectors and identification of the dark branch}

The eigenvectors of Eq.~\eqref{eq:H_pauli_revised} can be written as
\begin{equation}
\ket{\pm}=
\begin{pmatrix}
1\\
C_\pm
\end{pmatrix},
\qquad
C_\pm
=
-\frac{d_3}{d_1}
\pm
\sqrt{1+\left(\frac{d_3}{d_1}\right)^2}.
\label{eq:Cpm_revised}
\end{equation}
At $\Gamma$, one has $d_3=0$, so
\begin{equation}
C_\pm=\pm1.
\end{equation}
The two eigenstates are then the symmetric and antisymmetric combinations
\begin{equation}
\ket{+}\propto
\begin{pmatrix}
1\\
1
\end{pmatrix},
\qquad
\ket{-}\propto
\begin{pmatrix}
1\\
-1
\end{pmatrix}.
\label{eq:Gamma_eigenvectors}
\end{equation}

To identify the dark state, we note that at $\Gamma$ the TE polarization vectors satisfy $\vec{u}_{+1}=\vec{u}_y$ and $\vec{u}_{-1}=-\vec{u}_y$. The far-field amplitudes of the two basis states therefore cancel for the symmetric superposition $\ket{+}\propto(1,1)^T$, while they add for the antisymmetric superposition $\ket{-}\propto(1,-1)^T$. Accordingly, the dark BIC branch is the one continuously connected to $C=1$ at $\Gamma$. Expanding Eq.~\eqref{eq:Cpm_revised} and using Eq.~\eqref{eq:d1_leading} on this branch gives
\begin{equation}
C
\simeq
1-\frac{q_x}{n_g(U+i\gamma)}+O(q^2).
\label{eq:C_dark_revised}
\end{equation}

\subsection{Far-field electric field near $\Gamma$}

The in-plane radiated electric field is obtained by recombining the far-field contributions of the two basis guided modes with their respective TE polarization vectors,
\begin{align}
E_x
&\propto
\cos\theta_{+1}+C\cos\theta_{-1},
\label{eq:Ex_revised}\\
E_y
&\propto
\sin\theta_{+1}+C\sin\theta_{-1}.
\label{eq:Ey_revised}
\end{align}
Substituting the small-$\vec{q}$ expansions of Eq.~\eqref{eq:smallq_polarizations} and Eq.~\eqref{eq:C_dark_revised}, one finds
\begin{equation}
\mathbf{E}(\mathbf{q})
\propto
\begin{pmatrix}
-2q_y\\[2pt]
\dfrac{q_x}{n_g(U+i\gamma)}
\end{pmatrix}.
\label{eq:E_reduced_revised}
\end{equation}

This expression makes the origin of the two components transparent: the linear dependence on $q_y$ comes from the tilt of the TE polarization vectors away from $\pm\vec{u}_y$, whereas the linear dependence on $q_x$ arises from the correction to the eigenvector coefficients through the  denominator $U+i\gamma$.

\subsection{Connection to the winding number}

The far-field polarization orientation is encoded in the Stokes parameters
\begin{equation}
S_1=|E_x|^2-|E_y|^2,
\qquad
S_2=2\mathrm{Re}(E_xE_y^*).
\end{equation}
Using Eq.~\eqref{eq:E_reduced_revised}, the local polarization map near the BIC is governed by a linear transformation of $(q_x,q_y)$ whose orientation is controlled by
\begin{equation}
\mathrm{Re}\!\left(\frac{1}{U+i\gamma}\right)=\frac{U}{U^2+\gamma^2}.
\end{equation}
The Jacobian therefore changes sign when $U$ crosses zero, which yields
\begin{equation}
w_{\mathrm{BIC}}=-\mathrm{sgn}(U).
\label{eq:wBIC_revised}
\end{equation}

\section{Exceptional-point structure of the open-system band inversion}

In the main text, we refer to the dark/bright crossing as a ``band inversion'' of the radiative band structure. We expand here on what this means for an open (non-Hermitian) two-band system, since the gap closing relevant for the BIC winding flip is not the standard Hermitian level crossing.

The complex eigenvalues of the effective non-Hermitian Hamiltonian Eq.~\eqref{eq:H_full_revised} are $\Omega_\pm(\vec q)=d_0\pm\sqrt{d_1^2+d_3^2}$, with $d_1=U-i\gamma\cos\alpha$ and $d_3=q_x/n_g$ [Eq.~\eqref{eq:Omega_pm_revised}]. At normal incidence ($q_x=q_y=0$, $\cos\alpha=-1$), $d_3=0$ and $d_1=U+i\gamma$, so that $\Omega_+ - \Omega_- = 2(U+i\gamma)$. The real part of the splitting is therefore controlled by $U$, while the imaginary part is fixed by the radiative loss $\gamma$. Tuning $U$ through zero closes the \emph{real} part of the gap at $\Gamma$ but leaves a finite imaginary splitting: in the absence of $\Gamma$-point degeneracy, the two complex branches do not actually meet there.

\begin{figure}[!ht]
\begin{center}
	\includegraphics[width=0.8 \textwidth]{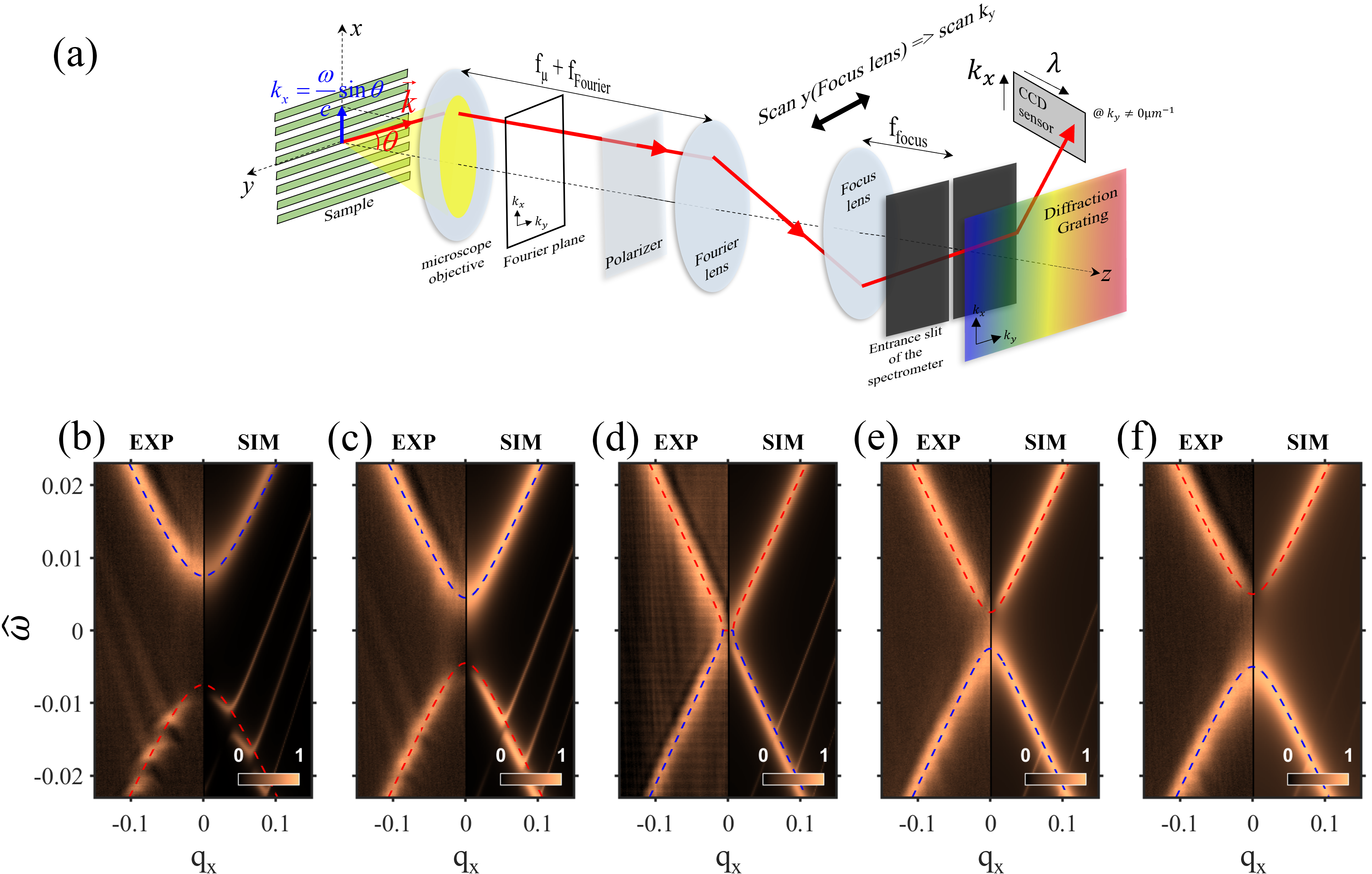}\\
\caption{\noindent\justifying{\textbf{Experimental setup and band-inversion mechanism}. (a) Sketch of the home-made Fourier set-up for the angle-resolved reflectivity measurements. The Fourier plane of the structure located at the focal length of the microscope objective is projected with the `Fourier' and `Focus' lenses to the entrance of the spectrometer. The slit of the spectrometer only selects the Fourier image in the $k_x$ direction for a given $k_y$, which can be controlled by the $y$ position of the `focus' lens. The signal is then dispersed by the spectrometer diffraction grating and collected by a CCD camera, resulting in a $k_x$-resolved measurement of the bands. (b)--(f) $k_x$-resolved contrast reflectivity maps of five different structures with different aspect ratios $w/a$ (diffractive coupling $U$): (b) $0.34$ ($-0.0075$), (c) $0.4$ ($-0.0045$), (d) $0.425$ ($0$), (e) $0.47$ ($0.0025$), (f) $0.505$ ($0.005$). For aspect ratios $w/a<0.44$ ($U<0$), the lower branch corresponds to the dark mode (BIC) and the upper branch to the bright mode. As the aspect ratio increases continuously, the dark and bright modes approach each other until they merge at two exceptional points (EPs) for $w/a\simeq 0.44$ ($U\simeq 0$, see preceding subsection). The two branches are swapped for $w/a>0.44$ ($U>0$), and the two modes then move apart as the aspect ratio increases further. This process is the non-Hermitian band-inversion mechanism. Panels (b) and (f) correspond to the two structures studied in the main article.}}
\label{fig:angle_resolved_setup}
\end{center}
\end{figure}

Where they do meet is at the \emph{exceptional points} (EPs) of the radicand $d_1^2+d_3^2$. Solving $d_1^2+d_3^2=0$ at $U=0$ gives $q_x = \pm i\,n_g\gamma$, which is purely imaginary and therefore unphysical; but in the experimentally relevant regime where the inversion parameter $U$ is swept across zero, the two real-momentum points $q_x = \pm n_g\sqrt{\gamma^2-U^2}$ (along the $q_x$ axis, for $|U|<\gamma$) become genuine EPs at which the eigenvalues and eigenvectors of $H$ coalesce. The aspect-ratio $w/a=0.44$ ($U=0$) sample shown in Fig.~\ref{fig:angle_resolved_setup}(d) is precisely the configuration in which these two EPs are positioned along the $q_x$ direction at the band edge.

The non-Hermitian band inversion is therefore a topological encirclement of an EP pair, not a single Hermitian crossing. The \emph{real-part} reordering of the dark and bright branches at $\Gamma$ across $U=0$ is what flips the eigenvector character on each branch (Eq.~\eqref{eq:Cpm_revised}), and is therefore the operationally relevant signature of the band inversion for the BIC winding flip established in the main text. The full open-system spectrum, including the EPs flanking the $\Gamma$ point, is shown in Fig.~\ref{fig:angle_resolved_setup} below.

\section{Numerical eigenmode visualization of the band-inversion transition}
\label{sec:COMSOL}

The angle-resolved reflectivity measurements of Fig.~\ref{fig:angle_resolved_setup} display the band-inversion mechanism through the merging and re-opening of two coupled resonances in the radiation continuum. The BIC itself, however, becomes intrinsically invisible to such a far-field measurement at the transition point $U=0$: its radiative width vanishes by definition, and its complex eigenfrequency degenerates with that of the bright partner at $\Gamma$, so the BIC cannot be singled out experimentally exactly at the inversion. To track the dark mode unambiguously across the transition --- and in particular to image the polarization-texture flip predicted in the body --- we complement the experiment with a finite-element eigenmode analysis carried out with COMSOL Multiphysics, which returns the complex eigenfrequencies $\Omega_\pm(\vec k)$ and the associated Bloch eigenmodes \emph{separately} for each branch, even when their complex spectra coincide.

\begin{figure}[!ht]
\begin{center}
	\includegraphics[width=0.95 \textwidth]{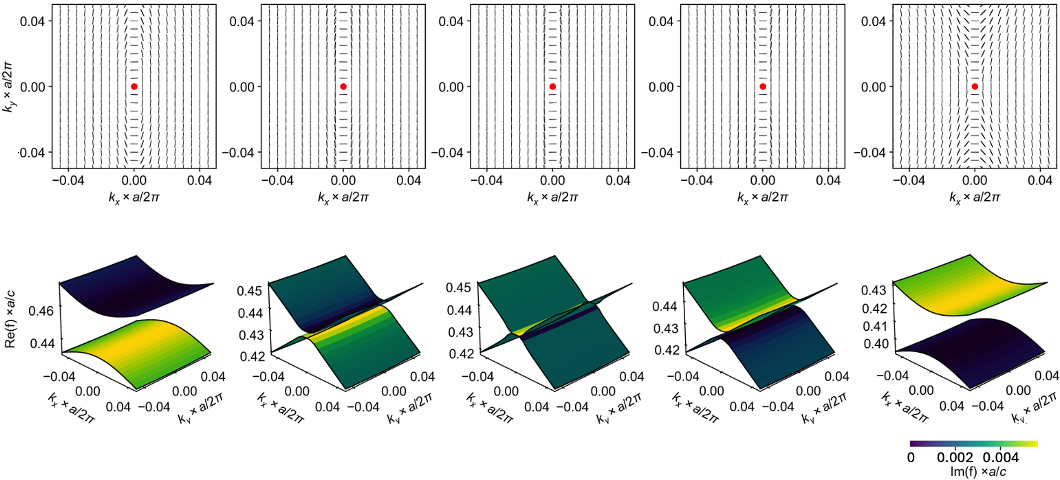}\\
\caption{\noindent\justifying{\textbf{Numerical eigenmode visualization of the BIC across the band inversion.} Finite-element eigenmode simulations (COMSOL Multiphysics) of a one-dimensional silicon grating ($\varepsilon_{\rm Si}=12$, thickness $100\,\mathrm{nm}$, etch depth $50\,\mathrm{nm}$, etch ratio $0.5$) of period $a=400\,\mathrm{nm}$ embedded in silica ($n_{\rm SiO_2}=1.45$), for five values of the grating-stripe width $w$ spanning the band-inversion transition: from left to right, $w=100$, $180$, $186.5$, $195$, and $300\,\mathrm{nm}$. \emph{Top row:} far-field polarization texture of the dark branch (BIC location marked by the red dot) in the reduced in-plane momentum plane $(k_x a/2\pi, k_y a/2\pi)$. The first-order vortex texture on either side of the transition collapses into a parallel ``shear'' pattern at the inversion ($w=186.5\,\mathrm{nm}$) and reappears with reversed handedness on the other side, in agreement with Eqs.~\eqref{eq:S_nabla_r_lorentzderiv} and \eqref{eq:wBIC_revised}. \emph{Bottom row:} real part of the complex eigenfrequencies $\mathrm{Re}\,\Omega_\pm(k_x,k_y)$ of the two coupled bands near $\Gamma$. The two bands approach each other, touch along the two real-momentum exceptional points predicted at $q_x=\pm n_g\gamma$ at the transition, and reopen with swapped dark/bright character, exposing the open-system, EP-mediated nature of the inversion that underlies the BIC winding flip established in the main text.}}
\label{fig:COMSOL_grating}
\end{center}
\end{figure}

We consider a one-dimensional silicon grating ($\varepsilon_{\rm Si}=12$) of period $a=400\,\mathrm{nm}$ and thickness $100\,\mathrm{nm}$ embedded in silica ($n_{\rm SiO_2}=1.45$), with an etch depth of $50\,\mathrm{nm}$ (etch ratio $0.5$). The structural parameter controlling the diffractive coupling $U$ in the analytical model of Eq.~\eqref{eq:H_full_revised} is the grating-stripe width $w$, which plays here the same role as the aspect ratio $w/a$ in Fig.~\ref{fig:angle_resolved_setup}. We have computed $\Omega_\pm(\vec k)$ and the associated far-field polarization texture for five values $w=100$, $180$, $186.5$, $195$, and $300\,\mathrm{nm}$ spanning the inversion. The result is shown in Fig.~\ref{fig:COMSOL_grating}.

The bottom row of Fig.~\ref{fig:COMSOL_grating} reports the real part of the two-band complex dispersion $\mathrm{Re}\,\Omega_\pm(k_x,k_y)$ around $\Gamma$. As $w$ is increased, the two bands approach each other at $\Gamma$, touch at the inversion ($w\simeq 186.5\,\mathrm{nm}$, central panel), and re-separate with their dark/bright character exchanged --- the non-Hermitian band-inversion sequence derived analytically in Eqs.~\eqref{eq:Cpm_revised}--\eqref{eq:Gamma_eigenvectors}. At the transition, the two real-eigenfrequency sheets meet along the two exceptional points predicted by Eq.~\eqref{eq:Omega_pm_revised} at $q_x=\pm n_g\,\gamma$ (along the $q_x$ axis), confirming that the gap closing relevant for the BIC winding flip is an EP-mediated, open-system inversion rather than a Hermitian level crossing.

The top row of Fig.~\ref{fig:COMSOL_grating} displays the far-field polarization texture of the dark branch in the reduced in-plane momentum plane. Far from the transition (leftmost and rightmost panels), the BIC carries a well-defined first-order vortex of opposite handedness on the two sides, $w_{\rm BIC}=\mp 1$, in agreement with Eq.~\eqref{eq:wBIC_revised}. As $w$ is tuned towards the inversion, the rotational character of the texture continuously fades and the polarization field collapses into a quasi-parallel ``shear'' pattern at $w=186.5\,\mathrm{nm}$. This is precisely the rank-deficient critical configuration anticipated by Eqs.~\eqref{eq:S_nabla_r_lorentzderiv} and \eqref{eq:w_from_Js}: at $\Delta_{\rm Re}=0$ the modulus gradient $\nabla_{\vec k}|r|^{\rm BIC}$ vanishes, the local Jacobian $J_s$ becomes singular, and the winding number is undefined. Beyond the inversion, the vortex re-emerges with reversed handedness. This eigenmode-resolved view therefore makes both signatures of the band inversion directly visible on the same set of structures: the EP pair flanking $\Gamma$ in the complex band structure, and the continuous flip of the BIC polarization texture from $w_{\rm BIC}=-1$ to $w_{\rm BIC}=+1$ through a non-vortex critical pattern at $U=0$.

\begin{figure}[!ht]
\begin{center}
	\includegraphics[width=0.8 \textwidth]{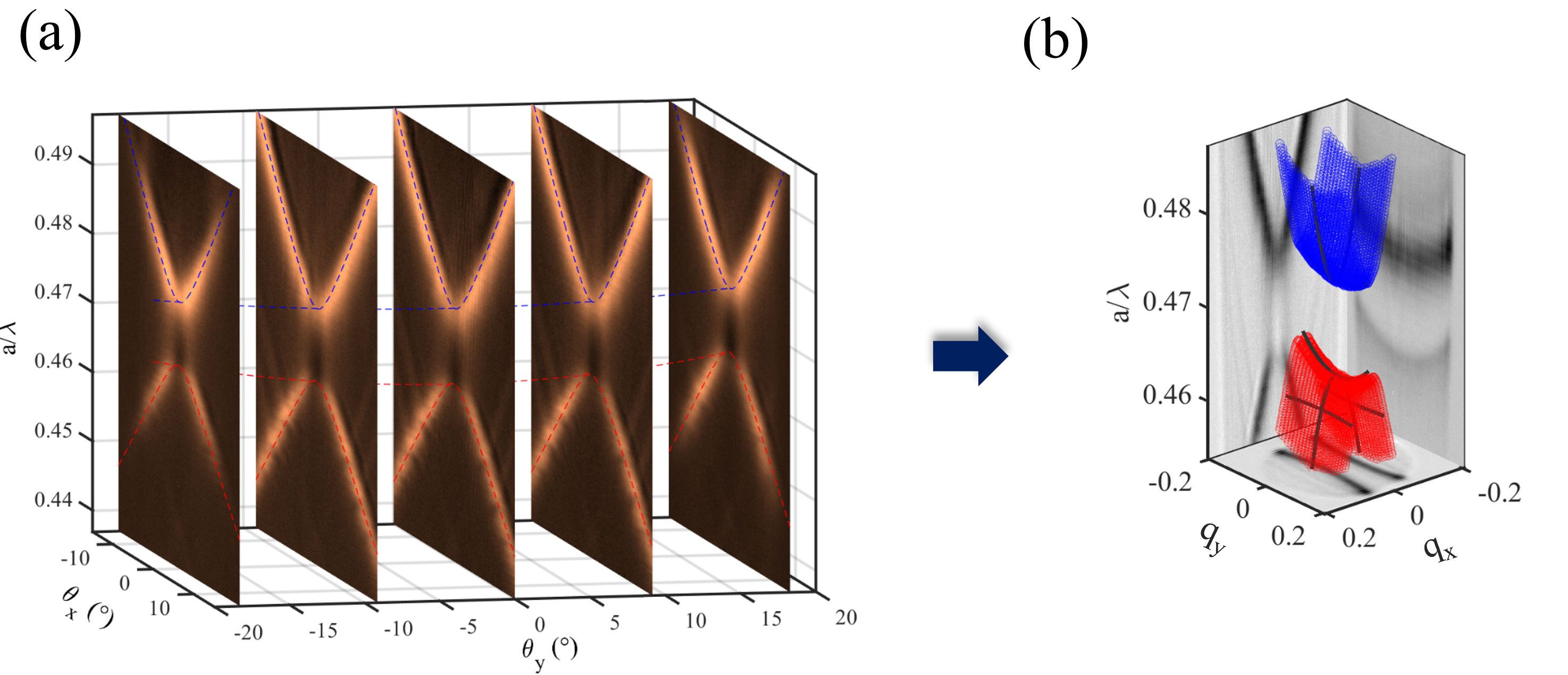} \\
\caption{\noindent\justifying{\textbf{ Tomography band reconstruction}  (a) Sketch illustrating the $k_x$-resolved measurements of the optical bands for different $k_y$ in the case of the structure with the diffractive coupling strength $U<0$. To perform tomographic bands reconstruction, 50 $k_x$-resolved measurements for $k_y$ varying from $-k_{max}$ to $k_{max}$ were performed for all the structures. Only five measurements are shown here for simplicity. For each $k_x$-resolved measurements at given $k_y$, each $k_x$ slices are fitted with two Lorentzian functions giving the modes wavelength $\lambda_m(k_x,k_y)$ (see blue and red dashed line), as well as their resonant amplitude $I_m(k_x,k_y)$. (b) Tomographic bands reconstruction of the structure with the diffractive coupling strength $U<0$. The projections in the momentum plane and in the ($k_x,a/\lambda$) plane are taken from the $k_y$-scan measurements presented in this figure. The projection in the ($k_y,a/\lambda$) plane is taken from additional measurements in which the sample was rotated by 90$^\circ$ to directly measure the $k_y$-resolved dispersion at $k_x$=0.}}
\label{fig:tomographic_illustration}
\end{center}
\end{figure}

\newpage

\begin{figure}[ht]
\begin{center}
	\includegraphics[width=0.8 \textwidth]{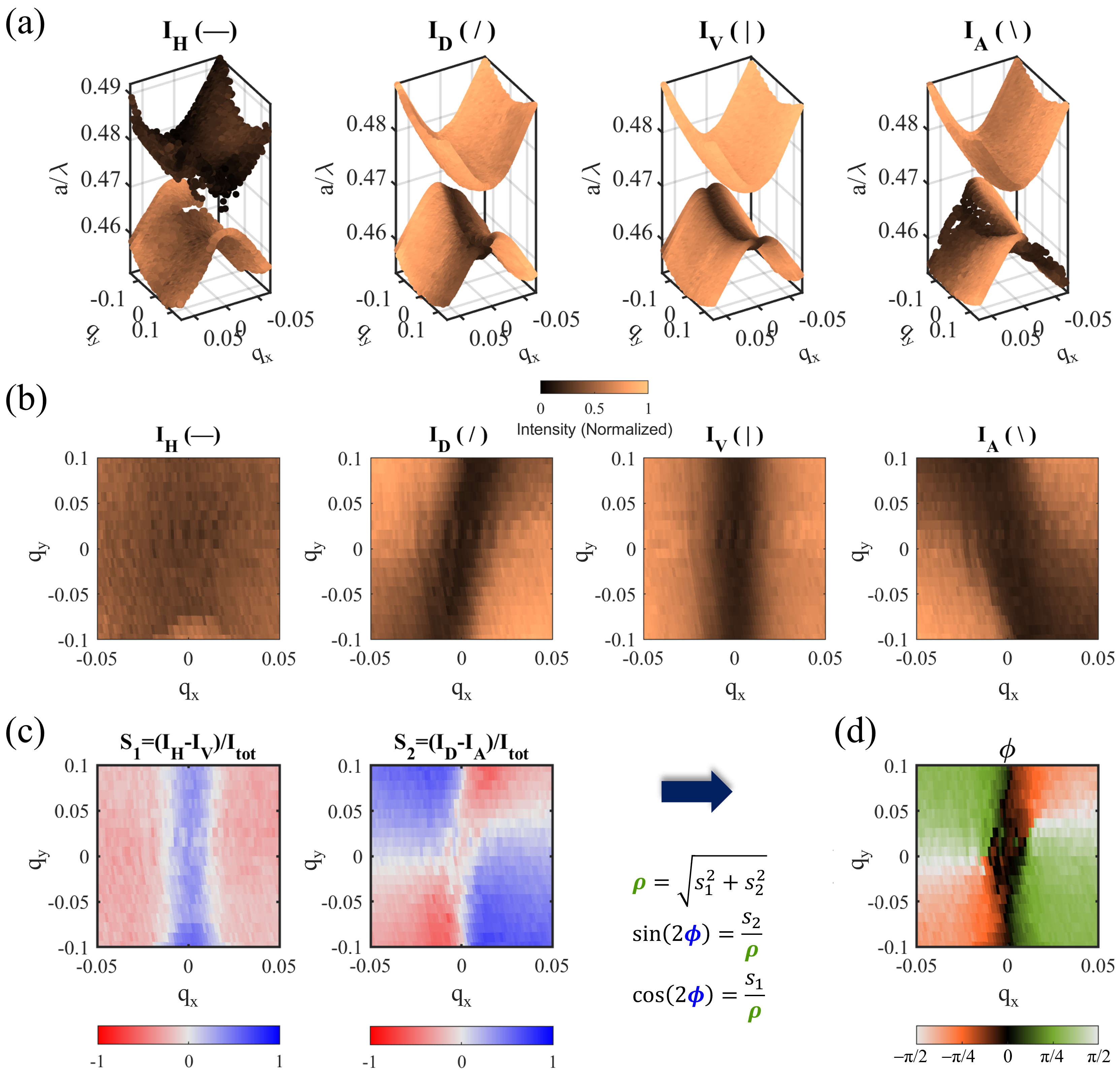}\\
\caption{\noindent\justifying{\textbf{Polarization orientation measurements} (a) Tomographic bands reconstruction of the structure with the diffractive coupling strength $U<0$ for four different linear polarizations: Horizontal, Diagonal, Vertical, and Anti-Diagonal, using polarization elements (polarizers, half-wave plates, quater-wave plates) introduced in the experimental set-up. The color-map texture indicates the intensity of the modes at each position in momentum space. A drop in intensity can be observed on the low-energy mode around the BIC at $q_x=q_y=0$ (b) Momentum-resolved intensity map of the low-energy mode in (a). (c) Momentum-resolved Stokes parameters $S_1=(I_H-I_V)/(I_H+I_V)$ and $S_2=(I_D-I_A)/(I_D+I_A)$ of the low-energy mode extracted from the intensity map in (b). (d) Farfield polarization orientation of the low-energy mode with a diffractive coupling $U<0$ calculated using the stokes parameters in (c). This method was used to measure the polarization orientation of all the modes studied in the article.}}\label{fig:polarization_orientation_setup}
\end{center}
\end{figure}

\newpage
\phantomsection


\end{document}